# Efficient Computation of Optimal Trading Strategies

Victor Boyarshinov
Dept Computer Science
Rensselaer Polytechnic Institute
Room 207, Lally Bldg
110 8th Street
Troy, NY 12180
boyarv@cs.rpi.edu

Malik Magdon-Ismail
Dept Computer Science
Rensselaer Polytechnic Institute
Room 207, Lally Bldg
110 8th Street
Troy, NY 12180
magdon@cs.rpi.edu

**Abstract**

Given the return series for a set of instruments, a *trading strategy* is a switching function that transfers wealth from one instrument to another at specified times. We present efficient algorithms for constructing (ex-post) trading strategies that are optimal with respect to the total return, the Sterling ratio and the Sharpe ratio. Such ex-post optimal strategies are useful analysis tools. They can be used to analyze the "profitability of a market" in terms of optimal trading; to develop benchmarks against which real trading can be compared; and, within an inductive framework, the optimal trades can be used to to teach learning systems (predictors) which are then used to identify future trading opportunities.

## 1 Introduction

A trader has in mind the task of developing a trading system that optimizes some profit criterion, the simplest being the total return. A more conservative approach is to optimize a risk adjusted return. Widely followed measures of risk adjusted returns are the Sterling Ratio and Sharpe Ratio. In an enviroment where markets exhibit frequent crashes and portfolios encounter sustained periods of losses, the Sterling ratio and the Sharpe ratio have emerged as the leading performance measures used in the industry. Given a set of instruments, a trading strategy is a switching function that transfers the wealth from one instrument to another. In this paper, we consider the problem of finding optimal trading strategies, i.e., trading strategies that maximize a given optimality criterion, on *historical data*. In particular, we consider optimal strategies with respect to the total cumulative return, as well as with respect to various risk adjusted measures of return (the Sterling ratio and variants of the Sharpe ratio). Finding the optimal trading strategy for non-zero transactions cost is a path dependent optimization problem even when the price time series is known. A brute force approach to solving this problem would search through the space of all possible trading strategies, keeping only the one satisfying the optimality criterion. Since the number of possible trading strategies grows exponentially with time, the brute force approach leads to an exponential time algorithm[1], which for all practical purposes is infeasible – even given the pace at which computing power grows. The contribution in this work is to give *efficient* (polynomial time) algorithms to compute the optimal trading strategy for various profit objectives, with or without constraints on the number of trades that can ber made. Our motivations for constructing such optimal strategies are:

(i) Knowing what the optimal trades are, one can take an inductive approach to real trading: on historical data, one can construct the optimal trades; one can then correlate various market and/or technical indicators with the optimal trades. These indicators can then be used to identify future trading opportunities. In a sense, one can try to *learn* to predict good trading opportunities based on indicators by emulating the optimal trading strategy. A host of such activity within the inductive framework, goes under the name of *financial engineering*.

(ii) The optimal trading performance under certain trading constraints can be used as a benchmark for real trading systems. For example, how good is a trading system that makes ten trades with a Sterling ratio of 4 over a given time period? One natural comparison is to benchmark this trading strategy against a Sterling-optimal trading strategy that makes at most ten trades over the same time period.

(iii) Optimal trading strategies (with or without constraints) can be used to quantitatively rank various markets (and time scales) with respect to their profitability according to a given criterion. So for example, one could determine the optimal time scale on which to trade a particular market, or given a set of markets, which is the most (risk adjusted) profit-friendly.

(iv) Given a stochastic model for the behavior of a pair of instruments, one can use the efficient algorithms presented here to construct *ex-ante* optimal strategies using simulation. To be more specific, note that the optimal strategy constructed by our algorithms requires full knowledge of the future price paths. The stochastic model can be used to generate sample paths for the instruments. These sample paths can be used to compute the optimal trading strategy given the current history and information set. One then has a sample set of future paths and corresponding optimal trading strategies on which to base the current action. Note that such a stochastic model for future prices would have to take into account correlations (including auto-correlations) among the instruments.

---

[1]The asymptotic running time of an algorithm is measured in terms of the input size $n$. If the input is a time sequence of $n$ price data points, then polynomial time algorithms have run time that is bounded by some polynomial in $n$. Exponential time algorithms have running time greater than some exponentially growing function in $n$ [6].

It is beyond the scope of the current discussion to develop these applications. Our main goal here is to present the algorithms for obtaining optimal trading strategies, *given* a price time series.

## 1.1 Trading Model

We now make the preceeding discussion more precise. We consider optimal trading strategies on two instruments, for concreteness, a stock $S$ and a bond $B$ with price histories $\{S_0, \ldots, S_n\}$ and $\{B_0, \ldots, B_n\}$ over $n$ consecutive time periods, $t_i, i \in \{1, \ldots, n\}$. Thus, for example, over time period $t_i$, the price of stock moved from $S_{i-1}$ to $S_i$. We denote the return sequence for the two instruments by $\{s_1, \ldots, s_n\}$ and $\{b_1, \ldots, b_n\}$ respectively: $s_i = \log \frac{S_i}{S_{i-1}}$, and correspondingly, $b_i = \log \frac{B_i}{B_{i-1}}$. We assume that one of the instruments is the benchmark instrument, and that all the equity is held in the benchmark instrument at the begining and end of trading. The bond is usually considered the benchmark instrument, and for illustration, we will follow this convention. The trivial trading strategy is to simply hold onto bond for the entire duration of the trading period. It is useful to define the excess return sequence for the stock, $\hat{s}_i = s_i - b_i$. When the benchmark instrument is the bond, the excess return as we defined it is the conventionally used one. However, one may want to measure performances of a trading strategy with respect to the S&P 500 as benchmark instrument, in which case the excess return would be determined relative to the S&P 500 return sequence. The excess return sequence for the bond is just the sequence of zeros, $\hat{b}_i = 0$. Conventionally, the performance of a strategy is measured relative to some trivial strategy, so the excess return sequence will be the basis of most of our performance measures.

**Definition 1.1 (Trading Strategy)** *A trading strategy $\mathcal{T}$ is a boolean n-dimensional vector indicating where the money is at the end of time period $t_i$:*

$$\mathcal{T}[i] = \begin{cases} 1 & \text{if money is in stock at the end of } t_i, \\ 0 & \text{if money is in bond at the end of } t_i. \end{cases}$$

*We assume that $\mathcal{T}[0] = \mathcal{T}[n] = 0$, i.e., all the money begins and ends in bond. A trade is entered at time $t_i$ if $\mathcal{T}[i] = 0$ and $\mathcal{T}[i+1] = 1$. A trade is exited at time $t_i$ if $\mathcal{T}[i] = 1$ and $\mathcal{T}[i+1] = 0$. The number of trades made by a trading strategy is equal to the number of trades that are entered.*

We make the following assumptions regarding the trading:

**A1** *[All or Nothing]*: The position at all times is either entirely bond or entirely stock.
**A2** *[No Market Impact]*: Trades can be placed without affecting the quoted price.
**A3** *[Fractional Market]*: Arbitrary amounts of stock or bond can be bougnt or sold at any time.
**A4** *[Long Strategies]*: We assume that we can only hold long positions in stock or bond.

Assumption **A1** is in fact not the case in many trading funds, for it does not allow legging into a trade, or holding positions in both instruments simultaneously. While this is technicaly a restriction, for many optimality criteria (for example return optimal strategies), one can show that there always exists an all-or-nothing optimal strategy. Thus, we maintain this simplifying assumption for our discussion. Further, such assumptions are typically made in the literature on optimal trading (see for example [14]). Assumptions **A2**–**A4** are rather mild and quite accurate in most liquid markets, for example foreign exchange. Assumption **A3** is needed for **A1**, since if all the money should be transfered to a stock position, this may necessitate the purchase of a fractional number of shares. Note that if $\mathcal{T}[i-1] \neq \mathcal{T}[i]$, then at the begining of time period $t_i$, the position was transferred from one instrument to another. Such a transfer will incur an instantaneous per unit transaction cost equal to the bid-ask spread of the instrument being transfered into. We assume that the bid-ask spread is some fraction ($f_B$ for bond and $f_S$ for stock) of the bid price.

We denote the equity curve for a trading strategy $\mathcal{T}$ by the vector $\mathcal{E}_\mathcal{T}$, i.e., $\mathcal{E}_\mathcal{T}[i]$ is the equity at the end of time period $t_i$, with $\mathcal{E}_\mathcal{T}[0] = 1$. Corresponding to the equity curve is the excess return sequence $r_\mathcal{T}$ for the trading strategy $\mathcal{T}$, i.e., for $i \geq 1$

$$r_\mathcal{T}[i] = \log \frac{\mathcal{E}_\mathcal{T}[i]}{\mathcal{E}_\mathcal{T}[i-1]} - b_i. \tag{1}$$

If we ignore the bid-ask spread, then the excess return in time period $t_i$ is given by

$$r_\mathcal{T}[i] = \hat{s}_i \mathcal{T}[i] = (s_i - b_i)\mathcal{T}[i]. \tag{2}$$

The bid-ask spread affects the return, reducing it by an amount depending on $\mathcal{T}[i-1]$. Denoting this transactions cost attributable to $\mathcal{T}[i]$ by $\Delta[i]$, we have that

$$\Delta[i] = -\mathcal{T}[i-1](1 - \mathcal{T}[i])\hat{f}_B - (1 - \mathcal{T}[i-1])\mathcal{T}[i]\hat{f}_S, \tag{3}$$

where $\hat{f}_S = \log(1+f_S)$ and $\hat{f}_B = \log(1+f_B)$. Thus, the bid-ask spread can be viewed as introducing an instantaneous return of $-\hat{f}_B$ or $-\hat{f}_S$ whenever the position is switched. To exactly which time period this transactions cost is applied may depend on the nature of the market, i.e., it may be applied to $r_\mathcal{T}[i]$, $r_\mathcal{T}[i-1]$ or $r_\mathcal{T}[i+1]$. The nature of the results will not change significantly for either of these options, so in our algorithms, we will generally make the choice that offers the greatest technical simplicity. For a trading strategy $\mathcal{T}$, we define the total return $\mu(\mathcal{T})$, the sum of the squared returns $s^2(\mathcal{T})$, the sum of squared deviations of the returns $\sigma^2(\mathcal{T})$ and the maximum drawdown $MDD(\mathcal{T})$ as follows,

$$\mu(\mathcal{T}) = \sum_{i=1}^{n} r_{\mathcal{T}}[i], \tag{4}$$

$$s^2(\mathcal{T}) = \sum_{i=1}^{n} r_{\mathcal{T}}[i]^2, \tag{5}$$

$$\sigma^2(\mathcal{T}) = \sum_{i=1}^{n} \left(r_{\mathcal{T}}[i] - \frac{1}{n}\mu(\mathcal{T})\right)^2 = s^2(\mathcal{T}) - \frac{1}{n}\mu^2(\mathcal{T}), \tag{6}$$

$$MDD(\mathcal{T}) = \max_{1 \leq k \leq l \leq n} -\sum_{i=k}^{l} r_{\mathcal{T}}[i]. \tag{7}$$

When it is clear from the context what trading strategy we are talking about, we will generally suppress the explicit dependence on $\mathcal{T}$. The performance measures that we consider in this paper are derived from these statistics. In particular, we are interested in the total return $\mu$, the Sterling ratio Strl, and variants of the Sharpe ratio, $\mathsf{Shrp}_1$ and $\mathsf{Shrp}_2$:

$$\mathsf{Strl}(\mathcal{T}) = \frac{\mu(\mathcal{T})}{MDD(\mathcal{T})}, \qquad \mathsf{Shrp}_1(\mathcal{T}) = \frac{\mu(\mathcal{T})}{\sigma(\mathcal{T})}, \qquad \mathsf{Shrp}_2(\mathcal{T}) = \frac{\mu(\mathcal{T})}{\sigma^2(\mathcal{T})}. \tag{8}$$

$\mathsf{Shrp}_1$ is the conventionally used Sharpe ratio. $\mathsf{Shrp}_2$ is a more risk averse performance measure, as it is more sensitive to the variance in the returns. Often, Strl as we have defined it is refered to as the Calmar ratio in the literature [8], and the Sterling ratio adds a constant (for example 10%) to the $MDD$ in the denominator [1]. Such a constant can easily be accomodated by our algorithms, and so we will maintain this simpler definition for the Sterling ratio.

The contribution of this paper is efficient algorithms for computing optimal trading strategies. We will use standard $O()$ notation in stating our results: let $n$ be the length of the returns sequences; we say that the run time of an algorithm is $O(f(n))$ if, for some constant $C$, the runtime is $\leq Cf(n)$ for any possible return sequences. If $f(n)$ is linear (quadratic), we say that the runtime is linear (quadratic). We will establish the following results.

**Theorem 1.2 (Return Optimal Trading Strategies)** *A total return optimal trading strategy can be computed in linear time. Specifically,*

i. **Unconstrained Trading.** *A trading strategy $\mathcal{T}_\mu$ can be computed in $O(n)$ such that for any other strategy $\mathcal{T}$, $\mu(\mathcal{T}_\mu) \geq \mu(\mathcal{T})$.*

ii. **Constrained Trading.** *A trading strategy $\mathcal{T}_\mu^K$ making at most $K$ trades can be computed in $O(K \cdot n)$ such that for any other strategy $\mathcal{T}^K$ making at most $K$ trades, $\mu(\mathcal{T}_\mu^K) \geq \mu(\mathcal{T}^K)$.*

**Proof:** See section 2. ∎

**Theorem 1.3 (Sterling Optimal Trading Strategies)** *A Sterling optimal trading strategy can be computed in near linear time. Specifically,*

i. **Unconstrained Trading.** *A trading strategy $\mathcal{T}_{\mathsf{Strl}}$ can be computed in $O(n \log n)$ such that for any other strategy $\mathcal{T}$, $\mathsf{Strl}(\mathcal{T}_{\mathsf{Strl}}) \geq \mathsf{Strl}(\mathcal{T})$.*

ii. **Constrained Trading.** *A trading strategy $\mathcal{T}_{\mathsf{Strl}}^{K}$ making at most $K$ trades can be computed in $O(n \log n)$ such that for any other strategy $\mathcal{T}^{K}$ making at most $K$ trades, $\mathsf{Strl}(\mathcal{T}_{\mathsf{Strl}}^{K}) \geq \mathsf{Strl}(\mathcal{T}^{K})$.*

**Proof:** See section 3. ∎

**Theorem 1.4 (Sharpe Optimal Trading Strategies)** *A Sharpe optimal trading strategy can be computed in near quadratic time. Specifically, trading strategies $\mathcal{T}_{\mathsf{Shrp}_1}$ and $\mathcal{T}_{\mathsf{Shrp}_2}$ can be found in $O(n^2 \log n)$ such that for any other strategy $\mathcal{T}$, $\mathsf{Shrp}_1(\mathcal{T}_{\mathsf{Shrp}_1}) \geq \mathsf{Shrp}_1(\mathcal{T})$ and $\mathsf{Shrp}_2(\mathcal{T}_{\mathsf{Shrp}_2}) \geq \mathsf{Shrp}_2(\mathcal{T})$*

**Proof:** See section 4. ∎

In all cases, our proofs are constructive, and so immediately give the algorithms for performing the desired computations. Next, we discuss the existing related work, followed by a detailed discussion of the algorithms, along with all necessary proofs.

**Related Work**

The body of literature on optimal trading is so enormous that we only highlight here some representative papers. The reasearch on optimal trading falls into two broad categories. The first group is on the more theoretical side where researchers assume that instrument prices satisfy some particular model, for example the prices are driven by a stochastic process of known form; the goal is to derive closed-form solutions for the optimal trading strategy, or a set of equations that the optimal strategy must follow. The main drawbacks of such theoretical approaches is that their prescriptions can only be useful to the extent that the assumed models are correct. Our work does not make any assumptions about the price dynamics to construct ex-post optimal trading strategies.

The second group of research which is more on the practical side is focused on exploring data driven / learning methods for the prediction of future stock prices moves and trading opportunities. Intelligent agents are designed by training on past data and their performance is compared with some benchmark strategies. Our results furnish (i) optimal strategies on which to train intelligent agents and (ii) benchmarks with which to compare their performance.

**Theoretical Approaches** Boyd et al. in [11] consider the problem of single-period portfolio optimization. They consider the maximization of the expected return subject to different types of constraints on the portfolio (margin, diversification, budget constraints and limits on variance or shortfall risk). Under certain assumptions on the returns distribution, they reduce the problem to numerical convex optimization. Similarily, Thompson in [14] considered the problem of maximizing the (expected) total cumulative return of a trading strategy under the assumption that the asset price satisfies a stochastic differential equation of the form $dS_t = dB_t + h(X_t)dt$, where $B_t$ is a Brownian motion, $h$ is a known function and $X_i$ is a Markov Chain independent of the Brownian motion. In this work, he assumes fixed transaction costs and imposes assumptions **A1, A2, A4** on the trading. He also imposes a stricter version of our assumption **A3**: at any time, the trader can have only 0 or 1 unit of stock. He proves that the optimal trading strategy is the solution of a free-boundary problem, gives explicit solutions for several functions $h$ and provides bounds on the transaction cost above which it is optimal never to buy the asset at all.

Pliska et al. in [4] and Bielecki et al. in [3] considered the problems of optimal investment with stochastic interest rates in simple economies of bonds and a single stock. They characterize the optimal trading strategy in terms of a nonlinear quasi-variational inequality and develop a numerical approaches to solving these equations.

Some work has been done within risk-return frameworks. Berkelaar and Kouwenberg in [2] considered asset allocation in a return versus downside-risk framework, with closed-form solutions for asset prices following geometric Brownian motions and constant interest rates. Liu in [10] consider the optimal investment policy of a constant absolute risk aversion (CARA) investor who faces fixed and proportional transaction costs when trading multiple uncorrelated risky assets.

Zakamouline in [16] studies the optimal portfolio selection problem using Markov chain approximation for a constant relative risk averse investor who faces fixed and proportional transaction costs and maximizes expected utility of the investor's end-of-period wealth. He identifies three disjoint regions (Buy, Sell and No-Transaction) to describe the optimal strategy.

Choi and Liu in [13] considered trading tasks faced by an autonomous trading agent. An autonomous trading agent works as follows. First, it observes the state of the environment. According to the environment state, the agent responds with an action, which in turn influences the current environment state. In the next time step, the agent receives a feedback (reward or penalty) from the environment and then perceives the next environment state. The optimal trading strategy for the agent was constructed in terms of the agent's expected utility (expected accumulated reward).

Cuoco et al. in [7] considered Value at Risk as a tool to measure and control the risk of the trading portfolio. The problem of a dynamically consistent optimal porfolio choice subject to the Value at Risk limits was formulated and they proved that the risk exposure of a trader subject to a Value at Risk limit is always lower than that of an unconstrained trader and that the probability of extreme losses is also decreased.

Mihatsch and Neuneier in [12] considered problem of optimization of a risk-sensitive expected return of a Markov Decision Problem. Based on an extended set of optimality equations, risk-sensitive versions of various well-known reinforcement learning algorithms were formulated and they showed that these algorithms converge with probability one under reasonable conditions.

**Data Driven Approaches**   Moody and Saffell in [9] presented methods for optimizing portfolios, asset allocations and trading systems based on a direct reinforcement approach, which views optimal trading as a stochastic control problem. They developed reccurent reinforcement learning to optimize risk-adjusted investment returns like the Sterling Ratio or Sharpe Ratio, while accounting for the effects of transaction costs.

Liu et al. in [15] proposed a learning-based trading strategy for portfolio management, which aims at maximizing the Sharpe Ratio by actively reallocating wealth among assets. The trading decision is formulated as a non-linear function of the latest realized asset returns, and the function cam be approximated by a neural network. In order to train the neural network, one requires a Sharpe-Optimal trading strategy to provide the supervised learning method with target values. In this work they used heuristic methods to obtain a locally Sharp-optimal trading strategy. The transaction cost was not taken into consideration. Our methods can be considerably useful in the determination of target trading strategies for such approaches.

## 2   Return-Optimal Trading Strategies

We use the notation $[t_i, t_j]$ to denote the set of time periods $\{t_i, t_{i+1}, \ldots, t_j\}$. In order to compute the return optimal strategies, we will use a dynamic programming approach to solve a more general problem. Specifically, we will construct the return optimal strategies for every prefix of the returns sequence. First we consider the case when there is no restriction on the number of trades, and then the case when the number of trades is constrained to be at most $K$. Although we maintain assumptions **A1-A4** for simplicity, **A1**, **A3** and **A4** can be relaxed without much additional effort.

### 2.1   Overview of the Algorithm

The basic idea of the algorithm is to consider the optimal strategy to time period $t_i$. This strategy must end in either stock or bond. Suppose that it ends in stock, then it must arrive at the final position in stock at $t_i$ by either passing through stock or bond at time $t_{i-1}$. Thus, the optimal strategy which ends in stock at time $t_i$ must be either the optimal strategy which passes through stock at time $t_{i-1}$ followed by holding the stock for one more time period, or the optimal strategy which passes through bond at time $t_{i-1}$ and then makes a trade into the stock for the next time period. Whichever is better among these two options yields the optimal strategy to time period $t_i$ that ends in stock. A similar argument applies to the optimal strategy to time $t_i$ that ends in bond. Thus, having computed the optimal strategies which end in stock and bond to time $t_{i-1}$, we can compute the optimal strategies which end in stock and bond to time $t_i$. This induction can be propagated to obtain the final result.

When there are constraints on the number of trades, we only need to slightly modify the above argument. We would like to compute the optimal strategy which ends (say) in stock and makes at most $K$ trades. Any such strategy has to be one of two possibilities: it makes at most $K$ trades ending in stock at time $t_{i-1}$, or it makes at most $K-1$ trades, ending in bond at time $t_{i-1}$. If it ended in bond, it can only make at most $K-1$ trades because one additional trade will be required to convert from bond at $t_{i-1}$ to stock at $t_i$. Thus the inductive construction will start with $K=0$ which is to hold bond. Assuming we have computed all the optimal strategies for $K=k$ to all times $\{t_i\}$, we can then compute all the optimal strategies for $K=k+1$ to all times.

We now present the details and proofs of correctness for these two algorithms.

## 2.2 Unconstrained Return-Optimal Trading Strategies

First we give the main definitions that we will need in the dynamic programming algorithm to compute the optimal strategy. Consider a return-optimal strategy for the first $m$ time periods, $[t_1, t_m]$. Define $\mathcal{S}[m,0]$ ($\mathcal{S}[m,1]$) to be a return-optimal strategy over the first $m$ periods ending in bond (stock) at time $t_m$. For $\ell \in \{0,1\}$, let $\mu[m,\ell]$ denote the return of $\mathcal{S}[m,\ell]$ over $[t_1, t_m]$, i.e., $\mu[m,\ell] = \mu(\mathcal{S}[m,\ell])$. Let $\text{PREV}[m,\ell]$ denote the penultimate position of the optimal strategy $\mathcal{S}[m,\ell]$ just before the final time period $t_m$.

The optimal strategy $\mathcal{S}[m,\ell]$ must pass through either bond or stock at time period $m-1$. Thus, $\mathcal{S}[m,\ell]$ must be the extension of one of the optimal strategies $\{\mathcal{S}[m-1,0], \mathcal{S}[m-1,1]\}$ by adding the position $\ell$ at time period $t_m$. More specifically,

$$S[m,\ell] = \begin{cases} \{S[m-1,1], \ell\}, \\ \{S[m-1,0], \ell\}. \end{cases}$$

In particular, $\mathcal{S}[m,\ell]$ will be the extension that yields the greatest total return. Using ẙq:ret and ẙq:tx, we have that

$$\mu(\{\mathcal{S}[m-1,0], \ell\}) = \mu[m-1,0] + \hat{s}_m \ell - \hat{f}_S \ell,$$
$$\mu(\{\mathcal{S}[m-1,1], \ell\}) = \mu[m-1,1] + \hat{s}_m \ell - \hat{f}_B(1-\ell).$$

Since $\mu[m, \ell]$ is the maximum of these two values, we have the following recursion,

$$\mu[m, \ell] = \max\left\{\mu[m-1, 0] + \hat{s}_m\ell - \hat{f}_S\ell,\ \mu[m-1, 1] + \hat{s}_m\ell - \hat{f}_B(1-\ell)\right\}.$$

The position of the optimal strategy $\mathcal{S}[m, \ell]$ just before time period $m$ is given by the ending position of the strategy that was extended. Thus,

$$\text{PREV}[m, \ell] = \begin{cases} 0 & \text{if } \mu[m-1, 0] + \hat{s}_m\ell - \hat{f}_S\ell \geq \mu[m-1, 1] + \hat{s}_m\ell - \hat{f}_B(1-\ell), \\ 1 & \text{otherwise.} \end{cases}$$

If we already know $\mu[m-1, 0]$ and $\mu[m-1, 1]$, then we can compute $\mu[m, \ell]$ and $\text{PREV}[m, \ell]$ for $\ell \in \{0, 1\}$ in constant time. Further, we have that $\mu[1, 0] = 0$ and $\mu[1, 1] = \hat{s}_1 - \hat{f}_S$, and so, by a straight forward induction, we can prove the following lemma.

**Lemma 2.1** $\text{PREV}[m, \ell]$ *for all* $\ell \in \{0, 1\}$ *and* $m \leq n$ *can be computed in* $O(n)$.

The optimal strategy $\mathcal{T}_\mu$ is exactly $\mathcal{S}[n, 0]$. $\text{PREV}[n, 0]$ gives the position at $t_{n-1}$, and the optimal way to reach $\text{PREV}[n, 0]$ at $t_{n-1}$ is given by optimal strategy $\mathcal{S}[n-1, \text{PREV}[n, 0]]$. Continuing backward in this fashion, it is easy to verify that we can reconstruct the full strategy $\mathcal{T}_\mu$ using the following backward recursion:

$$\begin{aligned} \mathcal{T}_\mu[n] &= 0, \\ \mathcal{T}_\mu[m] &= \text{PREV}[m+1, \mathcal{T}_\mu[m+1]], \text{ for } 1 \leq m < n. \end{aligned}$$

Thus, a single backward scan is all that is required to compute $\mathcal{T}_\mu[i]$ for all $i \in \{1, \ldots, n\}$, which is linear time, and so we have proved the first part of Theorem 1.2. Further, it is clear that the algorithm requires memory that is linear in $n$ to store $\text{PREV}[m, \ell]$. While we have assumed that the algorithm works with excess returns, the optimal strategy does not depend on this assumption, thus the algorithm works correctly even with the actual return sequences. The generalization of this algorithm to $N > 2$ instruments is straightforward by suitably generalizing a trading strategy. $\mathcal{S}[m, \ell]$ retains its definition, except now $\ell \in \{0, \ldots, N-1\}$. To compute $\mu[m, \ell]$ will need to take a maximum over $N$ terms depending on $\mu[m-1, \ell']$, and so the algirithm will have runtime $O(Nn)$.

One concern with the unconstrained optimal strategy is that it may make too many trades. It is thus useful to compute the optimal strategy that makes at most a given number of trades. We discuss this next.

## 2.3 Constrained Return-Optimal Strategies

We suppose that the number of trades is constrained to be at most $K$. It is more convenient to consider the number of jumps $k$, which we define as the sum of the number of trades entered and the number exited. For a valid trading strategy, the number of trades entered equals the number of trades exited, so $k = 2K$. Analogous to $\mathcal{S}[m, \ell]$ in the previous section, we define $\mathcal{S}[m, k, \ell]$ to be the optimal trading strategy to time period $t_m$ that makes at most $k$ jumps ending in instrument $\ell$. Let $\mu[m, k, \ell]$ be the return of strategy $\mathcal{S}[m, k, \ell]$, and let $\text{PREV}[m, k, l]$ store the pair $(k', \ell')$, where $\ell'$ is the penultimate position of $\mathcal{S}[m, k, \ell]$ at $t_{m-1}$ that leads to the end position $\ell$, and $k'$ is the number of jumps made by the optimal strategy to time period $t_{m-1}$ that was extended to $\mathcal{S}[m, k, \ell]$.

The algorithm once again follows from the observation that the the optimal strategy $\mathcal{S}[m, k, \ell]$ must pass through either bond or stock at $t_{m-1}$. A complication is that if the penultimate position is bond and $\ell = 0$, then at most $k$ jumps can be used to get to thhe penultimate position, however, if $\ell = 1$, then only at most $k - 1$ jumps may be used. Similarily if the penultimate position is stock. We thus get the following recursion,

$$\mu[m, k, 0] = \max\left\{\mu[m-1, k, 0],\ \mu[m-1, k-1, 1] - \hat{f}_B\right\},$$
$$\mu[m, k, 1] = \max\left\{\mu[m-1, k-1, 0] + \hat{s}_m - \hat{f}_S,\ \mu[m-1, k, 1] + \hat{s}_m\right\}.$$

This recursion is initialized with $\mu[m, 0, 0] = 0$ and $\mu[m, 0, 1] = \text{NULL}$ for $1 \leq m \leq n$. Once $\mu[m, k, \ell]$ is computed for all $m, \ell$, then the above recursion allows us to compute $\mu[m, k+1, \ell]$ for all $m, \ell$. Thus, the computation of $\mu[m, k, \ell]$ for $1 \leq m \leq n$, $0 \leq k \leq 2K$ and $\ell \in \{0, 1\}$ can be accomplished in $O(nK)$. Once again, the strategy that was extended gives $\text{PREV}[m, k, \ell]$,

$$\text{PREV}[m, k, 0] = \begin{cases} (k, 0) & \text{if } \mu[m-1, k, 0] > \mu[m-1, k-1, 1] - \hat{f}_B, \\ (k-1, 1) & \text{otherwise.} \end{cases}$$

$$\text{PREV}[m, k, 1] = \begin{cases} (k-1, 0) & \text{if } \mu[m-1, k-1, 0] + \hat{s}_m - \hat{f}_S > \mu[m-1, k, 1] + \hat{s}_m, \\ (k, 1) & \text{otherwise.} \end{cases}$$

Since computing $\mu[m, k, \ell]$ immediately gives $\text{PREV}[m, k, \ell]$, we have the following lemma,

**Lemma 2.2** $\text{PREV}[m, k, \ell]$ for all $\ell \in \{0, 1\}$, $m \leq n$ and $k \leq 2K$ can be computed in $O(nK)$.

$T_\mu^K$ is given by $S[n, 2K, 0]$, and the full strategy can be reconstructed in a single backward scan

using the following backward recursion (we introduce an auxilliary vector $\kappa$),

$$\begin{aligned}
\mathcal{T}_\mu^K[n] &= 0, \\
(\kappa[n-1], \mathcal{T}_\mu^K[n-1]) &= \text{PREV}[n, 2K, \mathcal{T}_\mu^K[n]], \\
(\kappa[m], \mathcal{T}_\mu^K[m]) &= \text{PREV}[m+1, \kappa[m+1], \mathcal{T}_\mu^K[m+1]], \text{ for } 1 \leq m < n-1.
\end{aligned}$$

Since the algorithm needs to store $\text{PREV}[m, k, \ell]$ for all $m, k$, the memory requirement is $O(nK)$. Once again, it is not hard to generalize this algorithm to work with $N$ instruments, and the resulting run time will be $O(nNK)$.

## 3 Sterling-Optimal Trading Strategies

It will first be useful to discuss some of the $MDD$ properties of the return-optimal strategy $\mathcal{T}_\mu$, as these properties will have implications on our algorithm to determine Sterling-optimal strategies. For a strategy $\mathcal{T}$, it is useful to define the cumulative return series, $C_\mathcal{T}[i]$ as the sum of the returns, $C_\mathcal{T}[i] = \sum_{j=1}^{i} r_\mathcal{T}[j]$. Note that $\mu(\mathcal{T}_\mu) = C_{\mathcal{T}_\mu}[n] \geq C_\mathcal{T}[n] = \mu(\mathcal{T})$ for any strategy $\mathcal{T}$. The equity curve is given by $\mathcal{E}_\mathcal{T}[i] = exp\left(C_\mathcal{T}[i] + \sum_{j=1}^{i} b_j\right)$.

### 3.1 Upper bound on the Maximum drawdown of a Sterling Optimal Strategy

First, we will upper bound $MDD(\mathcal{T}_\mu)$. Intuitively, the $MDD$ of the return optimal trading strategy can not be larger than the bid-ask spread. This bound will be useful because the $MDD$ of the $\mathcal{T}_\mu$ serves as an upper bound for the $MDD$ of the Sterling-optimal strategy,

**Lemma 3.1** $MDD(\mathcal{T}_{\text{Strl}}) \leq MDD(\mathcal{T}_\mu)$.

**Proof:** By definition, $\frac{\mu(\mathcal{T}_{\text{Strl}})}{MDD(\mathcal{T}_{\text{Strl}})} \geq \frac{\mu(\mathcal{T})}{MDD(\mathcal{T})}$ for any $\mathcal{T}$. Thus, $MDD(\mathcal{T}_{\text{Strl}}) \leq \frac{\mu(\mathcal{T}_{\text{Strl}})}{\mu(\mathcal{T})} MDD(\mathcal{T})$ for any $\mathcal{T}$. Choosing $\mathcal{T} = \mathcal{T}_\mu$ and noting that $\mu(\mathcal{T}_{\text{Strl}}) \leq \mu(\mathcal{T}_\mu)$, we obtain the desired result. ∎

Since the cost (in terms of the cumulative return) of entering and exiting a trade is $-(\hat{f}_S + \hat{f}_B)$, no segment of the optimal trading strategy $\mathcal{T}_\mu$ should lose more than this in return.

**Lemma 3.2** For any $i < j$, $C_{\mathcal{T}_\mu}[j] - C_{\mathcal{T}_\mu}[i] \geq -(\hat{f}_S + \hat{f}_B)$.

**Proof:** Suppose, for contradiction, that for some $i < j$, $C_{\mathcal{T}_\mu}[j] - C_{\mathcal{T}_\mu}[i] < -(\hat{f}_S + \hat{f}_B)$. By setting $\mathcal{T}_\mu[i+1], \ldots, \mathcal{T}_\mu[j]$ to be all equal to 0, it is easy to verify that the cumulative return of the strategy must increase, which contradicts the optimality of $\mathcal{T}_\mu$. ∎

For technical convenience, we will assume that the transactions cost when entering a trade is assigned to the time period prior to the entry, and the transactions cost when exiting a trade is assigned to the time period after the trade. Note that just prior to entering, the position is 0 and so it will now have a return of $-\hat{f}_S$, and just after exiting, the position is 0, and will now have a return of $-\hat{f}_B$.

Let $f_{sp} = \hat{f}_S + \hat{f}_B$. From Lemma 3.2, no segment of the optimal strategy can lose more than $f_{sp}$, and so this immediately gives an upper bound on $MDD(\mathcal{T}_\mu)$. For the trivial strategy that makes no trades, the $MDD$ is 0. If a strategy makes exactly one trade, then there is a drawdown of at least $\hat{f}_S$ at the begining, and of at least $\hat{f}_B$ at the end. If at least two trades are made, then there is a drawdown of at least $f_{sp}$ between the exit of one trade and the entry of another, and since the drawdown cannot exceed $f_{sp}$, the $MDD$ must therefore equal $f_{sp}$. We thus have the following lemma.

**Lemma 3.3 ($MDD$ of $\mathcal{T}_\mu$)** If $\mathcal{T}_\mu$ makes no trades, $MDD(\mathcal{T}_\mu) = 0$. If $\mathcal{T}_\mu$ makes one trade, $\max\{\hat{f}_S, \hat{f}_B\} \leq MDD(\mathcal{T}_\mu) \leq f_{sp}$. If $\mathcal{T}_\mu$ makes at least two trades, $MDD(\mathcal{T}_\mu) = f_{sp}$.

note that if we relax assumption **A1**, then by legging into a trade, it may be possible to decrease the drawdown, in which case Lemma 3.3 would no longer be valid. We are now ready to discuss the $O(n \log n)$ algorithms to obtain Sterling-optimal trading strategies. First we will consider unconstrained Sterling optimal strategies, and then we will require number of trades $\leq K$.

## 3.2 Overview of the Algorithm

Based on the results in the previous section, the main observation is that if the Sterling optimal strategy makes two or more trades, then its $MDD$ will be the bid-ask spread, and hence it must be the return optimal strategy. Thus we only need to consider the case when the Sterling optimal strategy makes exactly one trade. A useful observatioin is that no optimal strategy will exit a trade in the middle of a sequence of positive returns or enter a trade in the middle of a sequence of negative returns. Thus the return sequence can be contracted by combining any sequence of positive returns into a single return, ans similarly any sequence of negative returns. Possible entry points $a_i$ correspond to the beginning of a positive return and exit points the end of a sequence of positive returns.

The main task is to find the Sterling optimal strategy making exactly one trade. We show that this problem can be reduced to convex hull operations in 2-dimensions, and hence obtain an $O(n \log n)$ algorithm for constructing the optimal solution.

In order to construct the Sterling optimal strategy which makes at most $K$ trades, the basic idea is to start with the return optimal strategy whose trade intervals cannot be enlarged (maximal return optimal strategies). If this strategy makes at most $K$ trades, then we are back to the unconstrained case. Otherwise, we show that one can successively merge neighboring trades in the return optimal strategy to obtain a sterling optimal strategy. The main difficulty in the algorithm is to determine which trades to merge, and we will show that a greedy merging strategy is in fact optimal.

### 3.3 Unconstrained Sterling-Optimal Strategies

For a degenerate trading system with all returns equal to zero, we define its Sterling ratio as 1. The only trading system with a $MDD$ of 0 is a degenerate trading system, so with this definition, the Sterling ratio is defined for all possible trading systems. The computation of the Sterling-optimal trading system breaks down into three cases, according to the number of trades its makes:

**Sterling-optimal that makes zero trades.** Sterling Ratio is 1.

**Sterling-optimal that makes one trade.** The trading strategy contains a single interval of 1's.

**Sterling-optimal that makes at least two trades.** Any trading system that makes at least two trades has an $MDD \geq f_{sp}$. Since $MDD(\mathcal{T}_\mu) \leq f_{sp}$ (Lemma 3.3), $\mathcal{T}_\mu$ has the smallest $MDD$ among all such systems. Since it also has the highest total return, we conclude that if the Sterling-optimal system makes at least two trades, then $\mathcal{T}_{\mathsf{Strl}} = \mathcal{T}_\mu$.

The first case is trivially computed. The third case, i.e., the Sterling optimal strategy that makes at least two trades can be computed in linear time using the dynamic programming algorithm to compute $\mathcal{T}_\mu$. If we also compute the Sterling-optimal system that makes exactly one trade, then, we solve our problem by taking the case with the maximum Sterling ratio. We now focus on finding the trading strategy that makes only one trade and has greatest Sterling Ratio among all such strategies.

Let $\mathcal{T}$ be a strategy that makes exactly one trade. The trade interval is the interval of time periods, $[t_i, t_j]$ on which $\mathcal{T} = 1$, i.e., the trade interval is an interval of 1's in the trading strategy. An elementary algorithm that considers all the $O(n^2)$ possible trade intervals, picking the best is a quadratic time algorithm. The remainder of this section is devoted to providing an algorithm which computes such a strategy in $O(n \log n)$ time, which will complete the proof of the first part of Theorem 1.3. In fact the algorithm that we present is a much more general algorithm that computes the single interval that optimizes a general class of optimality criteria. This algorithm will be useful when we discuss the Sharpe-optimal strategy.

Consider a consecutive sequence of time periods $t_i, t_{i+1}, \ldots, t_{i+k}$ where $k \geq 1$, with all the excess returns non-negative and the last one positive, i.e., $\hat{s}_i \geq 0, \hat{s}_{i+1} \geq 0, \ldots, \hat{s}_{i+k} > 0$.

**Lemma 3.4** *Either the optimal single trade interval does not intersect these time periods, or an optimal single interval can be chosen to contain this interval.*

**Proof:** Suppose that $\mathcal{T}[i+j] = 1$ and $\mathcal{T}[i+j+1] = 0$ for some $0 \leq j < k$. Extend the trading interval by setting $\mathcal{T}[i+j+1] = 1, \ldots, \mathcal{T}[i+k] = 1$, which adds positive return, without increasing the $MDD$, contradicting the optimality of the original interval. On the other hand, suppose that $\mathcal{T}[i+j] = 1$ and $\mathcal{T}[i+j-1] = 0$ for some $0 < j \leq k$. Once again, by extending the trading interval, setting $\mathcal{T}[i] = 1, \ldots, \mathcal{T}[i+j] = 1$, we add non-negative returns, without increasing the $MDD$ hence this new interval is at least as good as the previous interval. ∎

A similar result holds for a sequence of consecutive negative time periods, $t_i, t_{i+1}, \ldots, t_{i+k}$ where $k \geq 1$, with $\hat{s}_i \leq 0, \hat{s}_{i+1} \leq 0, \ldots, \hat{s}_{i+k} < 0$. If an optimal trading interval only intersects part of these time periods, this intersection can be removed without decreasing the Sterling ratio. Thus, by Lemma 3.4, any sequence of time periods with all returns non-negative (non-positive) can be condensed into a single time period, $t'_i = t_i + \cdots + t_{i+k}$, with $\hat{s}'_i = \hat{s}_i + \cdots + \hat{s}_{i+k}$. Further, this operation can be performed in linear time on the entire excess return sequence, so from now on we assume without loss of generality that the excess return sequence consists of alternating time periods of strictly positive and negative excess returns. If $\hat{s}_i < 0$, then $t_i$ cannot be the first 1 of a trade, since by entering one time period later, we exclude only this negative return and do not increase the $MDD$. Similarily, it cannot be the last 1 of a trade.

**Lemma 3.5** *The first 1 and the last 1 of the optimal trade interval must occus at time periods $t_f$ and $t_l$ for which $\hat{s}_f > 0$ and $\hat{s}_l > 0$.*

The pictorial illustration of this lemma is given below where we show the cumulative return curve. The time *instants* $a_i$ are the possible entry points, and the time instants $b_i$ are the possible exit points.

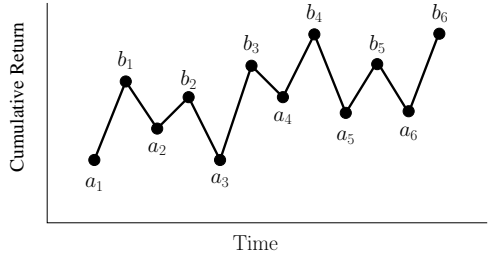

Let the alternating sequence of entry and exit points be $\{a_1, b_1, a_2, b_2, \ldots, a_k, b_k\}$ ($a_i$ are the entry points, and $b_i$ are the exit points). Note that after the preprocessing into alternating intervals, $k \leq \lceil n/2 \rceil$. Notice that without loss of generality, we can throw away the first interval if it has a negative return, as it can never be an entry point, and the last interval if it has a negative return, for a similar reason. The optimal trade interval will be of the form $(a_t, b_{t+r})$, $r \geq 0$.

Our algorithm for finding the Sterling-optimal interval will be to consider every possible starting point $a_i$, and find the Sterling-optimal interval with this point as starting point (i.e. we have to find the end point of this interval). As the algorithm proceeds, we keep track of the best entry point (and its corresponding exit point). The entry points $a_t$ are processed from right to left. After processing a new entry point $a_t$, we will modify the alternating sequence to facilitate faster processing of the remaining points. More specifically, we will delete the processed entry point and add a weight to the edge between $b_{t-1}$ and $b_t$ to preserve all the necessary information – we cannot simply delete the entry point $a_t$, since we have to keep track of maximum MDD that occurs in our activity interval. Since between $b_{t-1}$ and $a_t$ we have a drawdown of $b_{t-1} - a_t$, we need to keep this information in an edge weight connecting $b_{t-1}$ to $b_t$. At any stage of the algorithm, the edge weight connecting $b_{t-1}$ to $b_t$ will be equal to the MDD of the interval $[b_{t-1}, b_t]$, and this MDD is realized on a prefix of $[b_{t-1}, b_t]$, i.e. $MDD([b_{t-1}, b_t]) = C(b_{t-1}) - C(x)$, for some $x \in [b_{t-1}, b_t]$. This property is an invariant of the algorithm, which we will denote by $(*)$, and it will be maintained throughout the algorithm. We will explicitly show this weight attached to $b_t$ in parentheses as $(w_t)b_t$, where the value of $w_t$ appearing in parentheses indicates the weight.

We start our backward scan at the last entry point, $a_k$, for which there is only one possible interval $(a_k, b_k)$. We set the weight

$$w_k = b_{k-1} - a_k,$$

which is the drawdown of the interval from $b_{k-1}$ to $b_k$. We also store the current best interval

$$(a_k, b_k, \mathsf{Strl}_k),$$

where $\mathsf{Strl}_k = \frac{b_k - a_k - f_{sp}}{f}$; $f_{sp} = \hat{f}_S + \hat{f}_B$; and, $f = \max\{\hat{f}_S, \hat{f}_B\}$. We now delete the possible entry point $a_k$ from the sequence to give the processed sequence $\{a_1, b_1, ..., a_{k-1}, b_{k-1}, (w_k)b_k\}$. Note that $(a_k, b_k, \mathsf{Strl}_k)$ is a one-step trade, but we keep it here for simplicity. We now proceed to processing $a_{k-1}$ and so the process will continue until all $a_j$ have been processed.

In the general case, suppose we have processed (backwards) all the entry points up to (including) the entry point $a_{t+1}$, and are currently processing entry point $a_t$. The weighted exit sequence is

$$\{a_1, b_1, ..., a_t, b_t, (w_{t+1})b_{t+1}, \ldots, (w_{t+m})b_{t+m}\},$$

where $b_t, \ldots, b_{t+m}$ are the possible exit points – note that $t + m$ may not equal $k$ due to possible deletion of points which we discuss below. Assume that $\{b_{t+1} < \ldots < b_{t+m}\}$: this is true after we have processed the first start point (since the sequence consists only of one point), and we will maintain this condition by induction. If $b_t < b_{t+1}$, then the entire sequence of exit points is monotonically increasing. On the other hand, suppose that $b_t \geq b_{t+1}$. Then,

$$return(a_t, b_t) \geq return(a_t, b_{t+1}), \text{ and } MDD(a_t, b_t) \leq MDD(a_t, b_{t+1}),$$

and so $b_{t+1}$ need not be considered as an exit point for any optimal interval with entry point $a_t$ or earlier, because by stopping earlier at $b_t$, we do not decrease the cumulative return, nor increase the $MDD$. Hence, we can delete the possible exit point $b_{t+1}$. However, we must now update the weight in $(w_{t+2})b_{t+2}$ to store the new drawdown between $b_t$ and $b_{t+2}$, $w_{t+2} \leftarrow \max\{w_{t+1}, w_{t+2} + b_t - b_{t+1}\}$.

**Lemma 3.6** *If $b_t \geq b_{t+1}$, the weighted exit sequence is updated as follows:*

$$a_t, b_t, (w_{t+1})b_{t+1}, \ldots, (w_{t+m})b_{t+m} \to a_t, b_t, (\max\{w_{t+1}, w_{t+2} + b_t - b_{t+1}\})b_{t+2}, \ldots, (w_{t+m})b_{t+m}$$

The correctness of the weight updating rule above follows from the invariant (*) which is preserved by the transformation.

This process is continued until the next exit point after $b_t$ is either above $b_t$ or there are no remaining exit points after $b_t$. In either event, the new sequence of exit points available for $a_t$ is strictly monotonically increasing (by induction). Observe that any deletion of a possible exit point is a constant time operation. Further, since each deletion drops a point from the set $\{b_1, \ldots, b_k\}$, there can be at most $k - 1$ such deletions during the course of the *entire* algorithm. We thus have the following lemma.

**Lemma 3.7** *When $a_t$ is processed by the algorithm, the exit points $b_t < b_{t+1} < \cdots$ are monotonically increasing. The cost of maintaining this condition for the* entire *algorithm is $O(k)$ operations.*

When processing $a_t$, the weighted exit sequence is $\{a_1, b_1, \ldots, a_t, b_t, (w_{t+1})b_{t+1}, \ldots, (w_{t+m})b_{t+m}\}$. Suppose that $w_{t+2} < w_{t+3} < \ldots < w_{t+m}$. Initially this sequence is the empty sequence and so this condition holds, and once again, by induction we will ensure that this condition will always hold. Suppose that $w_{t+1} \geq w_{t+2}$. By construction, $b_{t+2} > b_{t+1}$ (Lemma 3.4), and so

$$return(a_t, b_{t+1}) < return(a_t, b_{t+2}), \text{ and } MDD(a_t, b_{t+1}) = MDD(a_t, b_{t+2}).$$

Thus, no optimal interval can have entry point $a_t$ (or earlier), and exit at $b_{t+1}$, because by exiting at $b_{t+2}$, the $MDD$ is not increased, however the total return is increased. Thus if $w_{t+1} \geq w_{t+2}$, we can remove the possible exit point $b_{t+1}$ from the weighted exit sequence and update $w_{t+2} \leftarrow w_{t+1}$. Note that this transformation also preserves the invariant (*).

**Lemma 3.8** *If $w_{t+1} \geq w_{t+2}$, the weighted exit sequence is updated as follows:*

$$a_t, b_t, (w_{t+1})b_{t+1}, \ldots, (w_{t+m})b_{t+m} \to a_t, b_t, (w_{t+1})b_{t+2}, \ldots, (w_{t+m})b_{t+m}$$

We continue removing exit points in this way until either there is only one weight left in the weighted exit sequence, or all the weights are strictly monotonically increasing (by induction).

Suppose that $w_{t+1} \leq f$. In this case, we observe that $b_t$ cannot be the exit of an optimal interval with entry $a_{t-r}$, where $r \geq 0$. To see this note that if $b_t - a_{t-r} - \hat{f}_S < 0$, then the return of this interval is negative and this interval cannot be an optimal interval. If $b_t - a_{t-r} - \hat{f}_S \geq 0$ then since the interval already has $MDD$ of at least $f$, so by continuing to $b_{t+1}$, we do not increase the $MDD$ but strictly increase the return, hence it cannot be optimal to exit at $b_t$.

**Lemma 3.9** *Let*

$$a_t, b_t, (w_{t+1})b_{t+1}, \ldots, (w_{t+m})b_{t+m}$$

*be the weighted exit sequence and let $T \in [t, \ldots, t+m]$ be such an index that $w_T < f$ and $w_{T_1} \geq f$. Then no Sterling-optimal interval that starts at $a_t$ or earlier can exit at $\{b_t, \ldots, b_{T-1}\}$.*

**Lemma 3.10** *When $a_t$ is processed by the algorithm, the weights $w_{t+1}$ are monotonically increasing. The cost of mainitaining this condition for the* entire *algorithm is $O(k)$ operations.*

We thus assume from now on that when processing entry point $a_t$, the weighted exit sequence $\{a_1, b_1, ..., a_t, b_t, (w_{t+1})b_{t+1}, \ldots, (w_{t+m})b_{t+m}\}$ with $m \geq 0$ satisfies the conditions of Lemmas 3.7 and 3.10. The first available exit gives the trade interval $(a_t, b_t)$. If $b_t - a_t - f_{sp} \leq 0$, i.e., if the return is not positive, then this cannot possibly be an optimal interval. Otherwise, the Sterling Ratio is

$$\mathsf{Strl}_t = \frac{b_t - a_t - f_{sp}}{f},$$

where $f_{sp} = \hat{f}_S + \hat{f}_B$ and $f = \max\{\hat{f}_S, \hat{f}_B\}$. Now consider the exit points $b_T$, where $T > t$ and suppose that $b_T - w_{T+1} < a_t$. In this case, for $r > 0$,

$$return(a_t, b_{T+r}) < return(a_{T+1}, b_{T+r}),$$

and so no trade with entry at $a_t$, exiting at $b_{T+r}$ can possibly be optimal, since we could always enter later at $b_T - w_{T+1}$, exit at $b_{T+r}$, and strictly increase the return without increasing the drawdown. We are thus done with processing the entry point $a_t$, and we can proceed to $a_{t-1}$ after updating weight $w_t$ and comparing $\frac{b_T - a_t - f_{sp}}{f}$ with the current champion. Similarly, if $b_{\bar{t}} - w_{\bar{t}+1} < a_t$ for some $\bar{t} \in [t, \ldots, T-1]$, we are done with the starting point $a_t$, and we can proceed to $a_{t-1}$ after updating weight $w_t$. We assume that at any stage of the algorithm we keep the value of $min_{\bar{t} \in [t,\ldots,T-1]} b_{\bar{t}} - w_{\bar{t}+1}$ and thus this check can be done in constant time for any given point $a_t$. Thus, without loss of generality, we can assume that $b_T - w_{T+1} \geq a_t$ and $b_{\bar{t}} - w_{\bar{t}+1} \geq a_t$ for all $\bar{t} \in [t, \ldots, T-1]$. Since $w_{T+1} \geq f$, we conclude that $b_T - a_t \geq f$. A trade, entering at $a_t$ and exiting at $b_{T+r}$, $r > 0$ has total return $b_{T+r} - a_t - f_{sp}$. The next lemma gives the $MDD$.

**Lemma 3.11** *Assume that $b_T - w_{T+1} \geq a_t$ and $b_{\bar{t}} - w_{\bar{t}+1} \geq a_t$ for all $\bar{t} \in [t, \ldots, T-1]$. The trade $(a_t, b_{T+r})$, $r > 0$, has $MDD = w_{T+r}$.*

**Proof:** The local maxima of the cumulative return sequence for this trade are $\{0, b_t - a_t - \hat{f}_S, \ldots, b_T - a_t - \hat{f}_S, \ldots, b_{T+r} - a_t - \hat{f}_S\}$. Since $b_{\bar{t}} - w_{\bar{t}+1} - a_t \geq 0\ \forall \bar{t} \in [t, \ldots, T-1]$, $MDD$ of the interval $[a_t, b_T]$ is equal to $\hat{f}_S$.

Since $b_T - a_t - \hat{f}_S \geq 0$ and since the sequence of exit points is strictly increasing, $MDD([a_t, b_{T+r}]) = max(MDD([a_t, b_T]), MDD([b_T, b_{T+r}])) = max(\hat{f}_S, MDD([b_T, b_{T+r}]), \hat{f}_B)$ where $\hat{f}_B$ is the draw down after the last point $b_{T+r}$.

Since the drawdown at any time in a trade is given by the the diffenence between the previous maximum and the current cumulative return, $MDD([b_T, b_{T+r}))$ is at most $\max_{i \in [1,r]} w_{t+i}$. Since the weights are monotonically increasing, we see that this drawdown $\leq w_{t+r}$, which is achieved in the interval $(b_{t+r-1}, b_{t+r})$. Since $w_{t+r} \geq f = max(\hat{f}_S, \hat{f}_B)\ \forall r > 0$, we conclude $MDD([a_t, b_{T+r}]) = w_{t+r}$. ∎

**Summary:** For entry point $a_t$, the sequence of exit points $b_{T+r}$, $r \in [0, m]$ have cumulative returns $c_r = b_{T+r} - a_t - f_{sp}$ and $MDD$'s $d_r = w_{T+r}$ for $r > 0$ and $d_0 = f$. The task is to maximize $c_r/d_r$ with respect to $r$. The sequences $\{c_r\}$ and $\{d_r\}$ are both strictly increasing. We now describe a general algorithm for performing such a maximization.

### 3.3.1 Maximizing $\dfrac{c_r}{d_r}$

Fix $t$ and consider the set of trades $(a_t, b_{t+r})$, $r \geq 0$. Corresponding to each trade is the pair $(d_r, c_r)$. As discussed in the previous section, by construction, the sequences $\{d_r\}$ and $\{c_r\}$ are strictly increasing. On the two dimensional plane, consider the set of points $P_m = \{(d_r, c_r)\}_{r=0}^{m}$ defined by this set of pairs. Let $\mathbf{p} = (0, 0)$. Then the slope of the line joining $\mathbf{p}$ to $(d_r, c_r)$ is exactly the Sterling ratio of the trade $(a_t, b_{t+r})$. Thus, finding the optimal trade is equivalent to finding the upper-touching point from $\mathbf{p}$ to the convex hull of the set of points in $P_m$ (see illustration below). We call $\mathbf{p}$ the source point.

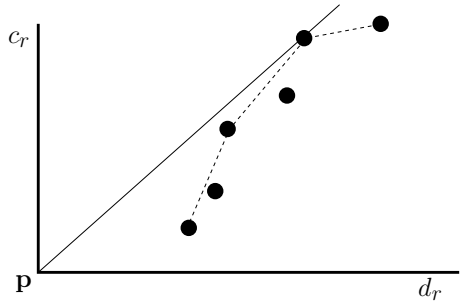

We get the same result if we define $P_m = \{(d_r, b_{t+r})\}_{r=0}^m$, $\mathbf{p} = (0, a_t + f_{sp})$. It is known that, given the convex hull, this touching line can found in time $O(\log A)$ where $A$ is number of the points on the convex hull, [5]. It is easy to see that $A \leq m + 1 \leq \lceil n/2 \rceil$. In order to compute this touching point in $O(\log A)$ time, one must be able to efficiently search through the convex hull. we now describe the mechanism for facilitating this search. The details are not important. What is important is that this data structure can be constructed and efficiently maintained as we update the set $\{d_r, c_r\}$. We will need to update this set as we process the points $a_t$ backwards.

We accomplish the fast construction of the upper tangent point of the convex hull by maintaining the convex hull as a doubly linked list, where each point in the convex hull maintains $O(\log A)$ pointers to other points in the convex hull. More specifically, for point $i$ we consider the set of points $\{(d_r, c_r)\}_{r=i}^m$ (point $i$, $(d_i, c_i)$ is the first point in this set). Corresponding to this set is some convex hull $C_i$. Point $i$ is also the first point in this convex hull (by the strict monotonicity of the $d_r, c_r$). Point $i$ maintains a forward pointer to points at position $2^j$ for $j \geq 1$ in the convex hull $C_i$ of the points $\{(d_r, c_r)\}_{r=i}^m$. Clearly point $i$ maintains $O(\log m)$ forward pointers. Each point also maintains backward pointers to any point that points forward to it. At point $j$, the backward pointers specific to the convex hull starting at point $i < j$ are maintained separately for each $i$ so that constant time access to this set of pointers is possible. An array of an array of pointers suffices to accomplish this. It is clear that the worst case memory requirement is $O(m \log m)$ pointers.

We now discuss the main operations we would like to be able to do on our set of points $\{(d_r, c_r)\}$ and the point $\mathbf{p}$ and still be able to update our convex hull data structure efficiently, and hence efficiently compute the upper tangent line. First we recall that the $d_r$ are monotonically increasing. Assume that each point $(d_r, c_r)$ stores all the necessary forward and backward pointers. Note that the point $(d_r, c_r)$ stores a forward pointer to the point $\mathsf{nxt}(r)$, which is the next point in the convex hull if all the points $d_0, \ldots, d_{r-1}$ were removed – remember that in this case, $d_r$ becomes leftmost, and so must be in the convex hull. We will see that these assumptions are maintained inductively. We state the next observation as a lemma.

**Lemma 3.12** *The initial convex hull with all the points is given by $(d_0, c_0)$ followed by the points pointed to by $\mathsf{nxt}(0), \mathsf{nxt}(\mathsf{nxt}(0)), \ldots$.*

As we move from $a_t$ to process $a_{t-1}$, the set of points $\{d_r, c_r\}$ will change. However, they do not change arbitrarily. As we will show, the changes can be reduced to combinations of the following four operations on the set of points $P_m$.

(i) Translate $\mathbf{p}$ by some given vector $\mathbf{v}$.

(ii) Translate all the points in $\{(d_r, c_r)\}$ by some given vector $\mathbf{v}$.

(iii) Remove the leftmost point $(d_0, c_0)$.

(iv) Add a new leftmost point $(d_{-1}, c_{-1})$.

If we can efficiently update our data structure after any of these four operations on the points in $P_m$, then we will be able to compute the upper tangent point efficiently for the new set. This is the content of the next lemma. Since it is a rather technical lemma, we give the proof in the appendix.

**Lemma 3.13** *Assuming that* **p**, $\{(d_r, c_r, \mathsf{nxt}_r)\}_{r=1}^m$ *are given, all the operations in (1)-(4) above can be accomplished in time $O(\log m)$. Further, in the event that a point is added or deleted, all necessary pointers are maintained.*

We will now show that these four operations are all that is needed for maximizing the Sterling ratio. Suppose that point $a_t$ has been processed in the algorithm – i.e., the upper tangent point (optimal trade with entry at $a_t$) from $\mathbf{p}_t = (0, a_t + f_{sp})$ to $P_m = \{(d_r, b_{t+r})\}_{r=0}^m$ has been computed. Now consider the addition $a_{t-1}$ to construct the new weighted exit sequence. First, delete the leftmost point $(d_0, b_T)$ from the convex hull (operation (iii)). Lets consider all posible operations that may take place when processinig $a_{t-1}$. There are several possibilities, corresponding to the cases which occur in the algorithm to ensure that the sequence $\{d_r, c_r\}$ is strictly increasing.

1. $b_{t-1} \geq b_t$. We remove (leftmost) points $b_{t+i}$ (operation (iii)), $i \geq 0$, until $b_{t-1} < b_{t+i+1}$, and the new weight $w'_{t+i+1}$ may have increased (Lemma 3.6). Deletion of points $b_{t+i}$ from the weighted exit sequence doesn't change the convex hull until $b_{t+i}$ is a possible exit point. After this point, deleting one point from the weighted exit sequence also deletes the corresponding leftmost point of the convex hull. At the very end of the sequence of deletions, we have to update the MDD of point $b_{t+i+1}$ from $w_{t+i+1}$ to $w'_{t+i+1}$, this can be done by deletion of the (leftmost) point $(w_{t+i+1}, b_{t+i+1})$ and addition of new (leftmost) point $(w'_{t+i+1}, b_{t+i+1})$. The total number of such removals during the entire algorithm is at most $n - 1$. When condition $b_{t-1} < b_t$ is satisfied, proceed to the next stage.

2. $b_{t-1} < b_t$.

   (i) $w_{t+1} \leq w_t$. We remove (leftmost) points $b_{t+i}$, $i \geq 0$, until $w_t < w_{t+i+1}$. Deletion of points $b_{t+i}$ from the weighted exit sequence doesn't change the convex hull until $b_{t+i}$ is a possible exit point. After this point, deleting one point from the weighted exit sequence also deletes the corresponding leftmost point of the convex hull. By Lemma 3.10, the total number of such removals cannot exceed $n - 1$ over the course of the entire algorithm. When condition $w_t < w_{t+1}$ is satisfied, proceed to the next stage.

   (ii) $b_{t-1} < b_t$ and $w_t < w_{t+1}$.

   (a) $f > w_t$. Add to the convex hull the point $(f, b_T)$.

(b) $f < w_t$. Add to the convex hull points $(w_t, b_t)$ and $(f, b_{t-1})$.

The new source point is $p_{t-1} = (0, a_{t-1} + f_{sp})$, which just corresponds to a shift of $\mathbf{p}_t$, and so once the new convex hull for the new weighted exit sequence is computed, an additional $O(\log n)$ operations are needed to find the new upper tangent point.

The total number of removals in the entire algorithm is $O(n)$. For each new entry point, we have at most a constant number of additions, and since the number of entry points is $O(n)$, we see that the total number of additions is $O(n)$. By Lemma 3.13, we have that each operation takes $O(\log n)$ in the worst case, thus the total run time is $O(n \log n)$. Collecting all the results together, we can find the Sterling-optimal strategies making zero, one or more than one trade in $O(n \log n)$ time, completing the proof of the first part of Theorem 1.3.

Note that by only maintaining exit points with weight at most some given constant, $MDD_0$, i.e., by truncating some region of the points to the right, this algorithm is easily extended to computing the Sterling-optimal strategy that uses exactly one trade and has an $MDD \leq MDD_0$.

**Proposition 3.14** *Given $MDD_0$, a Sterling-optimal strategy that uses exactly one trade and has $MDD \leq MDD_0$ can be computed in $O(n \log n)$ time.*

This result will be useful when we consider constrained Sterling-optimal strategies.

## 3.4 Constrained Sterling-Optimal Strategies

As with the return-optimal strategies, the unconstrained sterling-optimal strategies may make too many trades. Here, we consider the the Sterling-optimal strategy that makes at most $K$ trades, $\mathcal{T}_{\mathsf{Strl}}^K$. We refer to such strategies as $K$-Sterling-optimal. First, we present some properties of this strategy, before giving an efficient algorithm to compute it.

A maximal return-optimal strategy $\mathcal{T}_\mu^*$ is a return-optimal strategy whose trade intervals cannot be enlarged. Given any return-optimal strategy $\mathcal{T}_\mu$, in one (linear time) scan from left to right, we can enlarge any trade intervals maximally to the right as long as they keep the same return. Similarily, in one backward scan, we can extend all trade intervals maximally to the left. Since $\mathcal{T}_\mu$ can be computed in linear time, we conclude that

**Lemma 3.15** *A maximal return-optimal strategy $\mathcal{T}_\mu^*$ can be computed in linear time.*

If any trade interval of a maximal return-optimal strategy is extended in either direction, then the total return must strictly decrease. In the previous section, we gave an $O(n \log n)$ algorithm for computing the Sterling-optimal strategy with exactly 1 trade. We also saw that if the unconstrained Sterling-optimal strategy contains more than 1 trade, then it is $\mathcal{T}_\mu^*$. Fix $K$, and let the number of trades that $\mathcal{T}_\mu^*$ makes be $K_0 \leq K$. In this case $\mathcal{T}_{\mathsf{Strl}}^K = \mathcal{T}_\mu^*$, and we are done. Thus we only need to consider the case that $1 < K < K_0$. Some important properties of $\mathcal{T}_\mu^*$ are summarized below. When it is clear, We also use $\mathcal{T}_\mu^*$ to refer to the set of trading intervals $\{I_r\}_{r=1}^{K_0}$. Let $C_i = \sum_{j=1}^i \hat{s}_j$ denote the cumulative return sequence of the excess returns.

**Lemma 3.16** *Let $\mathcal{T}_\mu^*$ be maximal return-optimal. Let $I$ be an interval $[t_i, t_j]$.*

i. *If $I \in \mathcal{T}_\mu^*$, then, $\sum_{k=i}^j \hat{s}_k - f_{sp} \geq 0$ and $MDD(I) \leq f_{sp}$.*

ii. *Suppose $I$ does not intersect with any interval in $\mathcal{T}_\mu^*$ and let the return of $I$ ($\sum_{k=i}^j \hat{s}_k$) be $\mu(I)$. Then, $\mu(I) \leq f_{sp}$. If $I$ is adjacent to some interval of $\mathcal{T}_\mu^*$, then $\mu(I) < 0$. If $I$ is adjacent to two intervals in $\mathcal{T}_\mu^*$, then $\mu(I) < -f_{sp}$.*

iii. *Let $[t_l, t_r]$ and $[t_{l'}, t_{r'}]$ be two consecutive trade intervals in $\mathcal{T}_\mu^*$, $l \leq r < l' \leq r'$. Then, $C_r - C_{l'} > f_{sp}$, and for all $r < q < l'$, $C_{l'} < C_q < C_r$.*

Let $\{(a_i, b_i)\}$ denote the local minima and maxima of $\{C_i\}$, as in the previous section. Any trade of $\mathcal{T}_\mu^*$ or $\mathcal{T}_{\mathsf{Strl}}^K$ must enter (exit) at a local minimum (maximum). Further, the entry (exit) point must be a minimum (maximum) in the trade, otherwise we can shrink the trade, strictly increasing the return without increasing the $MDD$.

**Lemma 3.17** *Let $I = [t_l, t_r]$ be a trade interval of $\mathcal{T}_\mu^*$ or $\mathcal{T}_{\mathsf{Strl}}^K$. Then $C_l$ is a local minimum, $C_r$ is a local maximum, and for any $k$, with $l \leq k \leq r$, $C_l \leq C_k \leq C_r$*

We now give an important inclusion property of the strategy $\mathcal{T}_{\mathsf{Strl}}^K$. Essentially, it states that the Sterling optimal strategy can be constructed by merging some trades of the return optimal strategy. We give the technical proof in the appendix (which will be the case with most of the technical results in this section).

**Proposition 3.18** *Let $\mathcal{T}_\mu^*$ be a maximal return-optimal trading strategy. There exists a $K$-Sterling-optimal strategy $\mathcal{T}_{\mathsf{Strl}}^K$, $K > 1$, with the following property: if $I = [t_l, t_r]$ is any trading interval in $\mathcal{T}_{\mathsf{Strl}}^K$, then a prefix of $I$ and a suffix of $I$ are trades in the maximal return-optimal strategy $\mathcal{T}_\mu^*$.*

As a result of Proposition 3.18, we assume from now on that every interval of the sterling optimal strategy $\mathcal{T}_{\mathsf{Strl}}^K$ is prefixed and suffixed by (not necessarily distinct) intervals from a maximal return-optimal strategy that makes $K_0$ trades.

**Lemma 3.19** *If $1 < K \leq K_0$ then $\mathcal{T}_{\mathsf{Strl}}^K$ can be chosen to make exactly $K$ trades.*

Lemmas 3.18 and 3.19 indicate how $\mathcal{T}_{\mathsf{Strl}}^K$ can be constructed: start with all the intervals of a maximal return-optimal strategy $\mathcal{T}_\mu^*$ and then merge some neighbouring intervals, keeping the merging sequence that gave the best strategy. The number of possible merging sequences is exponential, however, we will now show that an efficient greedy merging algorithm gives the correct result.

Given two consecutive non-adjacent intervals $I_1 = [t_{l_1}, t_{r_1}]$, $I_2 = [t_{l_2}, t_{r_2}]$, where $I_1$ preceeds $I_2$, define the *bridge* $B(I_1, I_2) = [t_{r_1}, t_{l_2}]$ to be interval connecting $I_1$ with $I_2$. If $I_1$ and $I_2$ are intervals in a maximal return optimal strategy, then by Lemma 3.16, the $MDD$ of the bridge is $C_{r_1} - C_{l_2}$. Since $C_{r_1}$ is a maximum over the interval $[t_{l_1}, t_{l_2}]$, and $C_{l_2}$ is a minimum over the interval $[t_{r_1}, t_{r_2}]$, we have that the $MDD$ of the union of these three intervals, $[t_{l_1}, t_{r_2}]$ is given by $\max\{MDD(I_1), C_{r_1} - C_{l_2}, MDD(I_2)\}$.

For every bridge $B(I_1, I_2)$, define the *closure* $Cl(B(I_1, I_2))$ to be the *smallest* interval $J = [t_l, t_r]$, in the return sequence, satisfying the following three properties.

$Cl_1$. $C_l \leq C_m \leq C_r$ for $l \leq m \leq r$, i.e., $C_l$ is a minimum and $C_r$ is a maximum in $[t_l, t_r]$.

$Cl_2$. $I_1, I_2 \subset J$, i.e., $J$ contains both $I_1$ and $I_2$.

$Cl_3$. $J$ is prefixed and suffixed by intervals from a maximal return-optimal strategy $\mathcal{T}_\mu^*$.

Note that a bridge may not have closure. For example, if the two last intervals $I_{l-1}, I_l$ in $\mathcal{T}_\mu^*$ are such that such that the end point $I_l$ is below the end point of $I_{l-1}$, then $B(I_{l-1}, I_l)$ doesn't have a closure. The next lemma shows that if the closure $J$ for a bridge exists, then not only is it unique, but any other interval satisfying $Cl_1$ - $Cl_3$ contains $J$.

**Lemma 3.20** *For two intervals $I_1, I_2$, if $Cl(B(I_1, I_2))$ exists, then it is unique. Moreover, for any other interval $I$ satisfying $Cl_1$ - $Cl_3$, $Cl(B(I_1, I_2)) \subseteq I$.*

**Proof:** Let $J_1 = [t_{l_1}, t_{r_1}]$ and $J_2 = [t_{l_2}, t_{r_2}]$ satisfy $Cl_1$ - $Cl_3$. Without loss of generality, assume that $t_{l_1} \leq t_{l_2} < t_{r_1} \leq t_{r_2}$. By construction, $J_1 \cap J_2 = [t_{l_2}, t_{r_1}]$ satisfies $Cl_1$ - $Cl_3$. Now let $Cl(B(I_1, I_2))$ be the intersection of all intervals that satisfy $Cl_1$ - $Cl_3$, concluding the proof. ∎

Suppose that bridge $B$ and $B'$ are bridges in $\mathcal{T}_\mu^*$ and that $Cl(B)$ contains $B'$. Then $Cl(B)$ satisfies $Cl_1$ - $Cl_3$ with respect to $B'$ and hence $Cl(B)$ also contains $Cl(B')$.

**Lemma 3.21** *Let $B$ and $B'$ be bridges in $\mathcal{T}_\mu^*$. If $B' \subset Cl(B)$, then $Cl(B') \subset Cl(B)$.*

Any interval in $\mathcal{T}_{\mathsf{Strl}}^K$ containing bridge $B$ satisfies properties $Cl_1$ - $Cl_3$ (Lemma 3.17 & Proposition 3.18), immediately yielding the following proposition.

**Proposition 3.22** *Let $I \in \mathcal{T}_{\mathsf{Strl}}^K$ and let $B$ be a bridge in $\mathcal{T}_\mu^*$.*

  i. *If $B \subset I$, then $Cl(B) \subset I$.*

  ii. *If $B$ does not have a closure, then no K-Sterling-optimal strategy can contain $B$.*

  iii. *A K-Sterling-optimal strategy with more than one trading interval and no bridges of $\mathcal{T}_\mu^*$ has $MDD = f_{sp}$. If it contains one or more bridges $B_i$ of $\mathcal{T}_\mu^*$, then $MDD = \max_i MDD(Cl(B_i))$.*

*iv. The MDD of a K-Sterling-optimal strategy with more than one trading interval can be one of at most $T + 1$ possible values where $T$ is the number of bridges between the intervals of $\mathcal{T}_\mu^*$.*

**Proof:** (i) and (ii) are immediate. (iv) follows from (iii), thus we only need to prove (iii). Let $I \in \mathcal{T}_{\mathsf{Strl}}^K$ contain the consecutive bridges $B_1, \ldots, B_K$, and hence their closures. From (i), it is clear that $MDD(I) \geq \max_i MDD(\mathcal{Cl}(B_i))$. It is also clear that $I = \cup_i^K \mathcal{Cl}(B_i)$. We show, by strong induction on $K$, a more general statement than we need: suppose that $I = \cup_i^K \mathcal{Cl}(B_i)$, then $MDD(I) \leq \max_i MDD(\mathcal{Cl}(B_i))$. If $K = 1$ then $I = \mathcal{Cl}(B_1)$ and the result is trivial; suppose it is true for up to $K - 1$ consecutive bridges, $K > 1$, and suppose that $I$ is the union of $K$ closures of consecutive bridges. Consider the first closure $\mathcal{Cl}(B_1)$. Let $I = [t_l, t_r]$ and $\mathcal{Cl}(B_1) = [t_l, t_{r'}]$, $t_{r'} \leq t_r$. By definition of $\mathcal{Cl}(B_1)$, $C_{r'}$ is a maximum over $[t_l, t_{r'}]$. Thus, $MDD(I) = \max\{MDD(\mathcal{Cl}(B_1)), MDD([t_{r'}, t_r])\}$. If $r = r'$, then $I = \mathcal{Cl}(B_1)$ and we are done. If $r < r'$, then $t_{r'+1}$ is the begining of some bridge $B_\kappa$. Let $I' = \cup_{i=\kappa}^K \mathcal{Cl}(B_i)$. Then, $[t_{r'}, t_r] \subseteq I'$ and so $MDD([t_{r'}, t_r]) \leq MDD(I')$. But $I'$ is the union of at most $K - 1$ closures, so by the induction hypothesis, $MDD(I') \leq \max_{i \geq \kappa} MDD(\mathcal{Cl}(B_i))$, concluding the proof. ■

We will distinguish between four types of bridges. Let $I_1 = [t_{l_1}, t_{r_1}]$, $I_2 = [t_{l_2}, t_{r_2}]$ be consecutive intervals in $\mathcal{T}_\mu^*$. The bridge $B = B(I_1, I_2)$ can be one of four types:

*regular.*      $C_{l_1} \leq C_{l_2}$ and $C_{r_1} \leq C_{r_2}$, i.e., $\mathcal{Cl}(B) = [l_1, r_2]$.
*right irregular.*    $C_{l_1} \leq C_{l_2}$ and $C_{r_1} > C_{r_2}$, i.e., $\mathcal{Cl}(B)$ contains the next bridge.
*left irregular.*     $C_{l_1} > C_{l_2}$ and $C_{r_1} \leq C_{r_2}$, i.e., $\mathcal{Cl}(B)$ contains the previous bridge.
*irregular.*      $C_{r_1} > C_{r_2}$ and $C_{l_1} > C_{l_2}$, i.e., $\mathcal{Cl}(B)$ contains both the next and previous bridges.

We define the *weight* of the bridge $W(B(I_1, I_2))$ as follows:

$$W(B(I_1, I_2)) = \begin{cases} C_{r_1} - C_{l_2} & \text{if } B(I_1, I_2) \text{ is regular,} \\ C_{r_1} - C_{l_2} & \text{if } B(I_1, I_2) \text{ is left irregular and the previous bridge is right irregular.} \\ C_{r_1} - C_{l_2} & \text{if } B(I_1, I_2) \text{ is left irregular and the previous bridge is irregular.} \\ +\infty & \text{otherwise.} \end{cases}$$

The general idea behind our algorithm is to start with a maximal return-optimal strategy and greedily merge pairs of intervals or pair of bridges according to the bridge weight, keeping track of the best $K$ intervals each time. When no more merging can occur, because we are down to $K$ intervals or all the bridge weights are $\infty$, we return the best $K$ intervals we have seen so far. More precisely, let $\mathcal{T}_\mu^* = \{I_1, \ldots, I_{K_0}\}$ be a maximal return-optimal trading strategy making $K_0$ trades. We denote this pool of trade intervals by $P_0$, the *base pool*. From pool $P_i$, we obtain pool $P_{i+1}$ by a single merge according to the following rule. Let $B = B(I_1, I_2)$ be the bridge with smallest weight. If $B = \infty$, stop (pool $P_{i+1}$ does not exist). Otherwise, there are two cases.

i. *Regular merge*: if $B$ is regular, merge $B$ with $I_1$ and $I_2$ to get a larger interval $I_{new} = [t_{l_1}, t_{r_2}]$. We now update the status (type and weight) of any neighboring bridges as follows:

- Previous bridge changes from right-irregular to regular (update type and weight).
- Previous bridge $B'$ changes irregular to left-irregular (update type). If the bridge previous to $B'$ is right-irregular or irregular then update weight.
- Next bridge changes from irregular to right-irregular (update type).
- Next bridge changes from left-irregular to to regular (update type and weight).

ii. *Irregular merge*: if $B$ is left irregular, denoting the previous bridge $B^*$, merge the two bridges and the interval between them into one bigger bridge $B_{new} = B^* \cup I_1 \cup B$. The status of bridges other than $B$ have not changed. The status and weight of $B$ may need updating.

Intervals are formed only by regular merges, so it is easy to see that intervals resulting from this merging procedure begin at a minimum and end at a maximum. New bridges are formed by irregular merges, and the resulting bridge must begin at a maximum and end at a minimum. The only bridge weights that could change are those that had weights of $\infty$. In such an event the weight will drop, but not below the weight of the original bridge used in the merge that led to the update.

**Lemma 3.23** *Let bridge $B$ with weight $w$ be involved in a merge, and suppose that the weight of bridge $B'$ is updated in the process from $\infty$ to $u$. Then, $w < u$.*

**Proof:** There are two cases:

i. The bridge involved in the merge was regular, i.e., two consecutive intervals $I_1 = [t_{l_1}, t_{r_1}]$ and $I_2 = [t_{l_2}, t_{r_2}]$ are merged with their bridge $B_{12} = B(I_1, I_2)$ with $W(B_{12}) = w$. Let the preceeding interval be $I_0 = [t_{l_0}, t_{r_0}]$ and the following interval be $I_3 = [t_{l_3}, t_{r_3}]$, and let the preceeding and following bridges be $B_{01}$ and $B_{23}$ respectively. If $B_{01}$ obtained finite weight, it cannot be left irregular, as it would remain left irregular after the merge, and hence its weight would still be $\infty$. Thus, we need only consider $B_{01}$ right irregular or irregular (i.e., $C_{r_0} > C_{r_1}$). Its weight becomes $u = C_{r_0} - C_{l_1} > C_{r_1} - C_{l_1}$. Since $B_{12}$ is regular, $w = C_{r_1} - C_{l_2} < C_{r_1} - C_{l_1}$ and so $w < u$. If $B_{23}$ obtained finite weight, then it could not be right regular or irregular as it could not become regular or left irregular after the merge. Thus, we need only consider $B_{23}$ left irregular ($C_{l_2} > C_{l_3}$). Its weight becomes $u = C_{r_2} - C_{l_3} > C_{r_2} - C_{l_2}$. Since $B_{12}$ is regular, $C_{r_1} \leq C_{r_2}$, and so $u > C_{r_1} - C_{l_2} = w$.

ii. The bridge involved in the merge was left-irregular, i.e., $B_{12} = [t_{r_1}, t_{l_2}]$ is left irregular, and $B_{01} = [t_{r_0}, t_{l_1}]$ is either right-irregular or irregular (in both cases, $C_{r_0} > C_{r_1}$). Let $w = C_{r_1} - C_{l_2}$ be the weight of $B_{12}$. The merged bridge is $B = B_{01} I_1 B_{12}$. If $B$ has finite weight (i.e. it is either regular or left-irregular), then its new weight is $u = C_{r_0} - C_{l_2} > C_{r_1} - C_{l_2} = w$. If $B$ is left-irregular or irregular, then it does not change the weight of any other bridge. If, on the other hand, $B$ became right-irregular or irregular, then it could affect the weight of the following bridge $B_{23}$, if $B_{23}$ was left-irregular ($C_{l_2} > C_{l_3}$). In this case, the weight of $B_{23} = [t_{r_2}, t_{l_3}]$ becomes $v = C_{r_2} - C_{l_3} > C_{r_2} - C_{l_2}$. But since $B_{12}$ was left-irregular, $C_{r_2} \geq C_{r_1}$, and so $v > C_{r_1} - C_{l_2} = w$.

∎

The next lemma shows that if all the bridge weights become $\infty$, any further merged pairs of intervals can never be part of a $K$-Sterling-optimal strategy.

**Lemma 3.24** *If all the bridge weights in a pool of intervals are $\infty$, then any further merged pairs of intervals from this pool can never be part of a $K$-Sterling-optimal strategy.*

**Proof:** (Lemma 3.24). Let $P_r$ be pool obtained from $P_0$ by some sequence of merging intervals with bridges of finite weight, and suppose that all the bridges in $P_r$ have infinite weight. In particular, this means that none of the bridges in $P_r$ are regular. Denote the bridges by $B_1, \ldots, B_m$, and consider bridge $B_k$. If $B_k$ is right irregular or irregular, then all following bridges are either right irregular or irregular since all bridges have finite weight. If a trading interval contains $B_k$, it must contain $B_{k+1}$ (since $B_k$ is right irregular or irregular), and so by induction, it must contain all the following bridges (and their closures). But, the last bridge does not have a closure (as it is right irregular or irregular), a contradiction. If on the other hand, $B_k$ is left irregular, then all preceeding bridges are left irregular as all bridges have infinite weight. If a trading interval contains $B_k$, it must contain $B_{k-1}$ (since $B_k$ is left irregular), and so by induction, it must contain all the preceeding bridges (and their closures). But, the first bridge does not have a closure (as it is left irregular), a contradiction. We conclude that $B_k$ cannot be in any trading interval. ∎

Each merge decreases the number of intervals and number of bridges by one. If we merge down to pool $P_{K_0-K}$, we are left with exactly $K$ intervals. We will show that $\mathcal{T}_{\mathsf{Strl}}^K$ can be chosen to be the best $K$ trades (with respect to total return) in one of these pools. Specifically, define $\mathcal{T}_j^K$ to be the $K$ intervals in $P_j$ with the highest total return. We say that a strategy is *coarser* than pool $P_i$ if the strategy can be obtained by a sequence of merges of some (or all) of the intervals in $P_i$. Clearly, $\forall i$, $P_{i+1}$ is coarser than $P_i$, as $P_{i+1}$ is obtained from $P_i$ after a single merge. Note that for a strategy to be coarser than $P_i$, it need not contain every trade in $P_i$, however if it contains part of any trade in $P_i$, then it contains the entire trade. Next, we show that after a merge, the $MDD$ of the remaining intervals is equal to the weight of the bridge involved in the merging.

**Lemma 3.25** *If pool $P_i$, $i \geq 1$, was obtained from $P_{i-1}$ by a merge involving a bridge of weight $w$, then the MDD of any interval in $P_i$ is at most $w$. If the merge created a new interval (i.e., the bridge was regular), then the MDD of the new interval is equal to $w$.*

**Proof:** In pool $P_0$, since any bridge is adjacent to two intervals of $\mathcal{T}_\mu^*$, its weight is at least $f_{sp}$ (Lemma 3.16). Consider sequence of pools $P_0, P_1, \ldots, P_r$, where bridge $B_i$ with weight $W(B_i)$ was the minimum weight bridge involved in the merge that resulted in pool $P_i$ from from $P_{i-1}$. By Lemma 3.23 bridges weights are non-decreasing, i.e., $W(B_i) \leq W(B_{i+1})$.

We now use induction on the index $i$. For $i = 1$, from Lemma 3.16, every interval in $P_0$ has $MDD$ at most $f_{sp}$. If $P_1$ was obtained from $P_0$ by an irregular merge, then all intervals of $P_1$ are intervals of $P_0$, with $MDD$ at most $f_{sp}$. Since $W(B_1) \geq f_{sp}$, the claim holds. If the merge was regular, then the $MDD$ is $W(B_1) \geq f_{sp}$ and the $MDD$ of all other intervals is at most $f_{sp}$. Thus, the claim holds for $P_1$.

Suppose the claim holds for all $j < i$ and consider pool $P_i$ which was obtained from $P_{i-1}$ using a merge involving $B_i$. By the induction hypethesis, the $MDD$ of any interval from $P_{i-1}$ is at most $W(B_{i-1}) \leq W(B_i)$. If $P_i$ that was obtained by an irregular merge, every interval of $P_i$ is an interval of $P_{i-1}$ and thus has $MDD$ at most $W(B_{i-1}) \leq W(B_i)$. Suppose that $P_i$ was obtained by a regular merge – all intervals except the merged interval are intervals of $P_{i-1}$. Consider the $MDD$ of the new interval, which is obtained by the regular merge $I_1 \cup B_i \cup I_2$. Since new intervals are created only through regular merges, it is easy to see by induction that property $\mathcal{Cl}_1$ holds for all the intervals in $P_{i-1}$, in particular it holds for $I_1$ and $I_2$. Since $B_i$ was regular, the $MDD$ of the new interval is $\max(MDD(I_1), W(B_i), MDD(I_2))$. By the induction hypothesis, $MDD(I_1) \leq W(B_{i-1})$ and $MDD(I_2) \leq W(B_{i-1})$, thus, $\max(MDD(I_1), W(B_i), MDD(I_2)) = W(B_i)$. ∎

First, we show that if a $K$-Sterling-optimal strategy makes $K$ trades, all of which are contained in intervals of one of the pools $P_i$, then a $K$-Sterling-optimal strategy exists which is composed of the $K$ intervals with highest return in some pool $P_j$ with $j \leq i$.

**Lemma 3.26** *If $K$ subintervals of the intervals of pool $P_i$ are a $K$-Sterling-optimal strategy, then for some $j \leq i$, the $K$ intervals with highest return of pool $P_j$ are a $K$-Sterling-optimal strategy.*

**Proof:** If $P_i = P_0$, then the claim is trivial. Suppose that $i > 0$, and let $\mathcal{T} = \{I_1, \ldots, I_K\}$ be the $K$-Sterling-optimal strategy whose trades are all subintervals of intervals in $P_i$. Consider the set $\mathcal{B}$ of all bridges in $\mathcal{T}_\mu^*$ that are contained in $\mathcal{T}$, $\mathcal{B} = \{B_i\}_{i=1}^r$. We can assume that $\mathcal{B}$ is not empty because if it were, then $\mathcal{T}$ is composed of intervals in $\mathcal{T}_\mu^*$, in which case the top $K$ intervals (with respect to return) in $\mathcal{T}_\mu^*$ are clearly optimal. Since $P_i$ contains all the intervals in $\mathcal{T}$, $P_i$ contains all the bridges in $\mathcal{B}$. Thus, there must exist $j \leq i$ such that $P_j$ contains all the bridges in $\mathcal{B}$ and no pool $P_k$, with $k < j$ has this property, i.e., $P_j$ was obtained from the previous pool by a regular merge involving a bridge $B^*$ which must contain some bridge $B_l \in \mathcal{B}$. Let $I$ be the interval in $\mathcal{T}$ that contains $B_l$. Then, $I$ must contain the whole bridge $B^*$, since if $B^*$ is the result of irregular merges, one of which involved bridge $B_l$, then $B^* \subset \mathcal{C}l(B_l)$, and $\mathcal{C}l(B_l) \subseteq I$ (Proposition 3.22). Since $B \subset I$, $MDD(\mathcal{T}) \geq MDD(I) \geq W(B^*)$. By Lemma 3.25, since $B^*$ was the last bridge involved in a merge, the $MDD$ of every interval in $P_j$ is at most $W(B^*)$. Since every interval of $\mathcal{T}$ is a subinterval of some interval in $P_j$, we conclude that $\mathcal{T}$ is contained in at most $K$ intervals of $P_j$. Let $\mathcal{T}_j^K$ be the top $K$ intervals in $P_j$. Then, the return of $\mathcal{T}$ is at most the return of $\mathcal{T}_j^K$. Further, $MDD(\mathcal{T}_j^K) \leq W(B^*) \leq MDD(\mathcal{T})$, and so $\mathsf{Strl}(\mathcal{T}_j^K) \geq \mathsf{Strl}(\mathcal{T})$, concluding the proof. ∎

We are now ready to prove the main result, which will lead to the greedy algorithm for constructing a $K$-Sterling optimal strategy.

**Theorem 3.27** *Let $j^*$ be such that $\mathsf{Strl}(\mathcal{T}_{j^*}^K) \geq \mathsf{Strl}(\mathcal{T}_j^K)$, $\forall j$. Then $\mathcal{T}_{j^*}^K$ is $K$-Sterling optimal.*

**Proof:** Let $S_0^K$ be a K-Sterling-optimal strategy that makes $K$ trades – by Lemma 3.19, such a strategy must exist. If $S_0^K$ has the same Sterling ratio as the trading strategy composed of the $K$ most profitable trades in $P_0$, then we are done. If not, then we know from Proposition 3.18 that $S_0^K$ is coarser than $P_0$. We prove the following statement for all $k \geq 1$

> $Q(k)$: Suppose there exists a $K$-Sterling-optimal strategy $S_{k-1}^K$ that makes $K$ trades and is coarser than $P_{k-1}$. Then either $S_{k-1}^K$ is composed of $K$ intervals of $P_k$, or there exists a $K$-Sterling-optimal strategy $S_k^K$ that makes $K$ trades and is coarser than $P_k$.

We know that $Q(1)$ is true. Suppose that $Q(k)$ is true for all $k \geq 1$, we then prove the proposition as follows. By an easy induction, we have that if none of the $S_{j-1}^K$ are composed of $K$ intervals in $P_j$ for all $j \leq m$, then there is a $K$-Sterling-optimal strategy $S_m^K$ making exactly $K$ trades that is coarser than $P_m$. Suppose that we can construct a total of $\kappa + 1$ pools, $P_i$ for $0 \leq i \leq \kappa \leq K_0 - K$. If $\kappa < K_0 - K$ then all the bridge weights in $P_\kappa$ are infinite. If $\kappa = K_0 - K$, then any further merging leads to fewer than $K$ intervals. In both cases, there cannot exist a $K$-Sterling-optimal strategy that is coarser than $P_\kappa$. Therefore, for some $j^* \leq \kappa$, the $K$-Sterling-optimal strategy $S_{j^*-1}^K$ is composed of $K$ intervals of $P_{j^*}$. By Lemma 3.26, there is a $K$-Sterling-optimal strategy $T_{\mathsf{Strl}}^K$ that is composed of the top $K$ intervals of some pool $P_l$, where $l \leq j^*$.

What remains is to show that $Q(k)$ is true for all $k \geq 1$. Suppose that $S_{k-1}^K$ is coarser than $P_{k-1}$ and is not composed of $K$ intervals in $P_k$. We show that there exists $S_k^K$ that is coarser than $P_k$. Since $S_{k-1}^K$ is coarser than $P_{k-1}$, it contains at least one bridge $B$ in $P_{k-1}$ with finite weight (because if it contains an infinite weight bridge, then it either contains the preceeding or following bridge; this argument continues analogously to the proof of Lemma 3.24 until we include a bridge of finite weight). Let $I$ be the interval of $S_{k-1}^K$ that contains $B$, and let $I_l$ and $I_r$ be intervals in $P_{k-1}$ (which are subintervals of $I$) connected by $B$. Let $B^*$ be the bridge in $P_{k-1}$ with minimum weight that was involved in the merge to get $P_k$ from $P_{k-1}$, and let $I_l^*$ and $I_r^*$ be the intervals in $P_{k-1}$ connected by $B^*$. If $B^* = B$ then $S_{k-1}^K$ is also coarser than $P_k$ and we are done, so suppose $B^* \neq B$. There are two possibilities:

(i) $B^*$ *is a regular bridge.* If $S_{k-1}^K$ does not contain $I_l^*$ or $I_r^*$, then $S_{k-1}^K \cap (I_l^* \cup B^* \cup I_r^*) = \emptyset$ and thus $S_{k-1}^K$ itself is coarser than $P_k$, and can be chosen as $S_k^K$. Suppose that $S_{k-1}^K$ contains $I_l^*$ and not $I_r^*$ (similar argument if it contains $I_r^*$ and not $I_l^*$). Thus some interval $I' \in S_{k-1}^K$ has as a suffix $I_l^*$. Suppose we construct $S_k^K$ by replacing interval $I'$ by interval $I' \cup B^* \cup I_r^*$. $S_k^K$ is then coarser than $P_k$. Since $B^*$ is regular, the return of $I' \cup B^* \cup I_r^*$ is at least as big as the return of $I'$. $I_r^*$ is either an interval of $P_0$ or was obtained by merging some intervals of $P_0$ through bridges with weight at most $W(B^*)$ (Lemma 3.25), and so $MDD(I_r^*) \leq W(B^*)$. Since the maximum cumulative return for $I'$ is attained at its right endpoint (Lemma 3.17) and the left endpoint of $I_r^*$ is a minimum in $I_r^*$, we have that $MDD(I' \cup B^* \cup I_r^*) = \max\{MDD(I'), W(B^*), MDD(I_r^*)\} = \max\{MDD(I'), W(B^*)\}$. Since $W(B^*) \leq W(B)$, we conclude that $MDD(S_k^K) \leq MDD(S_{k-1}^K)$, and thus $\mathsf{Strl}(S_k^K) \geq \mathsf{Strl}(S_{k-1}^K)$, which means that $S_k^K$ is also $K$-Sterling-Optimal. Finally, suppose that $S_{k-1}^K$ contains both $I_l^*$ and $I_r^*$, and consider the strategy $S_k^K$ obtained from $S_{k-1}^K$ by removing bridge $B$ and adding bridge $B^*$. $\mu(S_k^K) = \mu(S_{k-1}^K) + W(B) - W(B^*) \geq \mu(S_{k-1}^K)$. Since $W(B) \geq W(B^*)$, the $MDD$ cannot have increased, and so $S_k^K$ is $K$-Sterling-Optimal and coarser than $P_k$.

(ii) $B^*$ *is an irregular bridge.* Since $B^* = B(I_l^*, I_r^*)$ has finite weight, we can conclude that $B^*$ is left-irregular and the previous bridge $B_- = B(I_{l-1}^*, I_l^*)$ is right-irregular or irregular. Since $S_{k-1}^K$ does not contain $B^*$, by Lemma 3.17, there are two possibilities: $S_{k-1}^K$ does not contain $I_l^*$, in which case it also does not contain bridge $B_-$ and so $B_-$ and $B^*$ can be merged into one bridge without influencing $S_{k-1}^K$, i.e., $S_{k-1}^K$ is also more coarse than $P_k$; or, $S_{k-1}^K$ contains $I_l^*$ as one of its intervals. In this case, since $B^*$ is left-irregular, $\mu(I_l^*) < W(B^*) \leq W(B)$, and so by dropping $I_l^*$ from $S_{k-1}^K$ and breaking $I$ into two subintervals by removing $B$ from $I$ results in a profit increase of $W(B) - \mu(I_l^*) > 0$. Further, the $MDD$ cannot increase, so the new strategy makes $K$ trades and has strictly greater Sterling ratio than $S_{k-1}^K$, which contradicts optimality of $S_{k-1}^K$. Thus, $S_{k-1}^K$ cannot contain $I_l^*$ as one of its intervals.

Thus $Q(k)$ holds for all $k \geq 1$, concluding the proof. ∎

We are now ready to give the $O(n \log n)$ algorithm that establishes Theorem 1.3. First, we can compute the optimal strategy that makes only one trade in $O(n \log n)$ (Section 3.3), and compare this with the trivial strategy that makes no trades. It remains to compute the $K$-Sterling-optimal strategy and pick the best. We show how to do this in $O(n \log n)$ time.

First we obtain $\mathcal{T}_\mu^*$ in linear time. Suppose $\mathcal{T}_\mu^*$ makes $K_0 > K$ trades (as otherwise $\mathcal{T}_\mu^*$ is our solution). By Theorem 3.27, we only need to construct the pools $P_0, P_1, \ldots$, maintaining the pool with the optimal Sterling ratio for its top $K$ trades, as explained in the following algorithm.

1: Set $i = 0$; Sort (in decreasing order) the intervals from $P_0$ according to profit; Sort all the bridges with finite weight in increasing order. Let $B_i$ be the minimum weight bridge in $P_i$; Let strategy $S_i$ consist of the top $K$ intervals, and let $\mathsf{Strl}_{opt} = \mathsf{Strl}(S_i)$;
2: **while** $P_i$ contains at least $K$ intervals and at least one finite weight bridge **do**
3:     **if** $B_i = B(I_l, I_r)$ is regular **then**
4:         Regular merge to obtain $P_{i+1}$: remove $I_l, I_r, B_i$ from the interval and bridge orderings, and add back $I = I_l \cup B_i \cup I_r$ into the interval ordering; compute $\mu(I)$ and $MDD(I)$;
5:         Update neighboring bridge weigths and re-insert them back into the bridge ordering.
6:     **else if** $B_i = B(I_l, I_r)$ is left-regular **then**
7:         Irregular merge to obtain $P_{i+1}$: Let $B_-$ be the bridge left of $B_i$; remove $I_l, B_-, B_i$ from the interval and bridge orderings. Create the new bridge $B = B_- \cup I_l \cup B_i$, compute $W(B)$ and insert $B$ into the bridge ordering (note that $W(B)$ may be $\infty$).
8:     **end if**
9:     $i \leftarrow i + 1$; update $\mathsf{Strl}_i$; if $\mathsf{Strl}_{opt} < \mathsf{Strl}_i$, then $\mathsf{Strl}_{opt} \leftarrow \mathsf{Strl}_i$.
10: **end while**

The correctness of the algorithm follows from Theorem 3.27. We now analyse the run time of an efficient implementation of the algorithm. $P_0$ contains at most $n$ intervals and bridges. Each execution of the while loop reduces loop number of bridges and intervals by 1 each, so the while loop is executed at most $n/2$ times. Merging two intervals is a constant time operation. The profit of a new interval is the profit of the merged intervals minus the weight of the merging bridge (also computable in constant time). The $MDD$ of a new interval is the maximum of the $MDD$ of the merged intervals and the weight of the merging bridge (also computable in constant time). The weight of a new bridge takes constant time to compute, and updating the weights of the neighbour bridges is a constant time operation provided that pointers are maintained to them. These pointers can be updated in constant time as well. Thus the run time within the while loop is dominated by inserting into the bridge or interval orderings. At most a constant number of such such inserts into a *sorted list* need to be done, and each is an $O(\log n)$ operation [6]. To efficiently implement step 9, we maintain two sorted lists of the top $K$ intervals in the algorithm, sorted according to return and $MDD$. These can be maintained in $O(\log K)$ operations. The first allows us to update the total return of the top $K$ intervals in constant time, and the second allows us to update the $MDD$ of the top $K$ intervals (by taking the interval with largest $MDD$) in constant time. Thus the total running time of the while loop is $O(n \log n + n \log K) = O(n \log n)$ The preprocessing (step 1) is $O(n)$, and so does not contribute to the asymptotic run time.

# 4 Sharpe Optimal Trading Strategies

Another popular measure of the portfolio's risk-adjusted return is the Sharp Ratio. For trading strategy $\mathcal{T}$, we consider two versions of the Sharpe ratio, $\mathsf{Shrp}_1$ and $\mathsf{Shrp}_2$.

$$\mathsf{Shrp}_1(\mathcal{T}) = \frac{\mu(\mathcal{T})}{\sigma(\mathcal{T})}, \qquad \mathsf{Shrp}_2(\mathcal{T}) = \frac{\mu(\mathcal{T})}{\sigma^2(\mathcal{T})}. \tag{9}$$

Note that $\mathsf{Shrp}_2$ is more conservative in that it penalizes large variances more heavily. We introduce a simplified Sharpe ratio (SSR) $\mathsf{S}$ that will be instrumental to finding the optimal strategies,

$$\mathsf{S} = \frac{\mu}{s^2}.$$

It is easy to check that maximizing $\mathsf{Shrp}_1$ is equivalent to maximizing $\frac{\mu^2}{s^2}$, and that $\mathsf{Shrp}_2$ is given by $\frac{\bar{r}}{\frac{1}{n}s^2 - \bar{r}^2}$, where $\bar{r}$ is the mean return. We will relate the maximization of $\mathsf{Shrp}_1$ and $\mathsf{Shrp}_2$ to the maximization of $\mathsf{S}$.

Let $\mathcal{T}$ be a trading strategy that makes $K$ trades, with trading intervals $I_1, \ldots, I_K$. Each trade contributes a transactions cost of $-f_{sp}$ to the return sequence. In general, a trade contributes $\hat{f}_S^2 + \hat{f}_B^2$ to $s^2$. However, we will assume that $f_{sp} \ll 1$ and so we will ignore the contribution of the transactions cost to $s^2$. Alternatively, we can justify this by assuming that the transactions cost is spread finely over many time intervals. The sum over these small time intervals is finite, equal to $-f_{sp}$, however, the sum of squares over these small time intervals can be made arbitrarily small. Define the total return and sum of squared returns for each trading interval,

$$\mu_i = \mu(I_i) = \sum_{j \in I_i} r[j], \qquad s_i^2 = s^2(I_i) = \sum_{j \in I_i} r[j]^2.$$

We define $A_i$ as the contribution of trade $i$ to the mean return, and $B_i$ as the contribution of trade $i$ to the mean squared return (ignoring the effect of the transactions cost), i.e., $A_i = \frac{1}{n}(\mu_i - f_{sp})$ and $B_i = \frac{1}{n}s_i^2$. We define $A(\mathcal{T}) = \sum_{k=1}^{K} A_i$ (note that $\bar{r} = A(\mathcal{T})$) and $B(\mathcal{T}) = \sum_{k=1}^{K} B_i$ (note that $\frac{1}{n}s^2 = B(\mathcal{T})$).

## 4.1 Maximizing the Simplified Sharpe Ratio $S$

We will need the following technical lemma, which can be proved by an easy induction.

**Lemma 4.1** *Let $\mathcal{F} = \{\frac{a_1}{b_1}, \frac{a_2}{b_2}, \ldots, \frac{a_k}{b_k}\}$ be any set of fractions satisfying $b_i > 0$ and $\frac{c}{d} \leq \frac{a_i}{b_i} \leq \frac{a}{b}$, for all $i$, where $b, d > 0$. Then, $\frac{c}{d} \leq \frac{a_1 + a_2 + \ldots + a_k}{b_1 + b_2 + \ldots + b_k} \leq \frac{a}{b}$. The upper (resp. lower) bound is strict if at least one of the fractions in $\mathcal{F}$ is strictly upper (resp. lower) bounded by $\frac{a}{b}$ (resp $\frac{c}{d}$).*

Let $\mathcal{T}^*$ be an SSR-optimal strategy making $K > 1$ trades with trading intervals $I_1, \ldots, I_K$.

$$\mathsf{S}(\mathcal{T}^*) = \frac{\sum_{i=1}^{K} A_i}{\sum_{i=1}^{K} B_i} = \frac{A(\mathcal{T}^*)}{B(\mathcal{T}^*)},$$

**Lemma 4.2** $\frac{A_i}{B_i}$ *is a constant for every interval $i$, i.e., every trade is equivalent.*

**Proof:** Suppose that $\min_i \frac{A_i}{B_i} < \frac{A_j}{B_j}$ for some $j$ (strict inequality), and without loss of generality, assume that the minimum is attained for interval $I_1$. By Lemma 4.1, if we remove $I_1$, we get that

$$\mathsf{S}(I_1 \cup \cdots \cup I_K) = \frac{\sum_{i=1}^{K} A_i}{\sum_{i=1}^{K} B_i} < \frac{\sum_{i=2}^{K} A_i}{\sum_{i=2}^{K} B_i} = \mathsf{S}(I_2 \cup \cdots \cup I_K),$$

which contradicts the optimality of $\mathcal{T}^*$ implying that $\min_i \frac{A_i}{B_i} = \frac{A_j}{B_j}$ for all $j$. ∎

**Corollary 4.3** *An SSR-optimal trading strategy making one trade exists.*

**Proposition 4.4** *An SSR-optimal strategy making one trade can be found in $O(n \log n)$ time.*

**Proof:** By Corollary 4.3, we are guaranteed the existence of such a strategy. It suffices to find the single interval $I$ maximizing $\sum_{i \in I} r[i] / \sum_{i \in I} r[i]^2$. Consider all intervals starting at position $i$ and define $c_k = \sum_{j=i}^{k} r[j]$ and $d_k = \sum_{j=i}^{k} r[j]^2$. We wish to find $k$ to maximize $c_k/d_k$. If we have done this for position $i$, we now consider position $i - 1$. We show that the algorithm in Section 3.3.1 can be used. Trade intervals starting at $i - 1$ correspond to shifting all the $c_k$ by $r[i-1]$, and all the $d_k$ by $r[i-1]^2$. Both these operations simply correspond to shifting the origin point $\mathbf{p}$ to $\mathbf{p}' = \mathbf{p} - (r[i-1], r[i-1]^2)$. We then add a new leftmost point at $\mathbf{p}$. Since each update takes $O(\log n)$, and the optimal interval for the new points can be found in $O(\log n)$, the entire algorithm runs in $O(n \log n)$. ∎

## 4.2 Maximizing $\mathsf{Shrp}_2$

Ignoring the $f_{sp}^2$ term in the denominator changes the denominator slightly, so we introduce the slightly different quantity $\overline{\mathsf{Shrp}_2}$. Specifically,

$$\mathsf{Shrp}_2(\mathcal{T}) = \frac{A(\mathcal{T})}{\frac{d}{n} f_{sp}^2 + B(\mathcal{T}) - A^2(\mathcal{T})}, \qquad \overline{\mathsf{Shrp}_2}(\mathcal{T}) = \frac{A(\mathcal{T})}{B(\mathcal{T}) - A^2(\mathcal{T})},$$

where $d$ is the number of trades in $\mathcal{T}$. By the Cauchy-Schwarz inequality, for any trading strategy, $\sum r[i]^2 \geq \frac{1}{n}(\sum_i r[i])^2$. Since we are only interested in strategies for which $A(\mathcal{T}) \geq 0$, we have **Lemma 4.5** *For strategy $\mathcal{T}$, if $A(\mathcal{T}) > 0$ then $B(\mathcal{T}) - A^2(\mathcal{T}) > 0$*

We will show that maximizing $\overline{\mathsf{Shrp}}_2$ is closely related to a constrained optimization of the SSR, and that maximizing $\overline{\mathsf{Shrp}}_2$ is not too far from maximizing $\mathsf{Shrp}_2$.

Let $\mathcal{T}^*_\mu$ be return optimal, with return $\mu(\mathcal{T}^*_\mu) = \mu^*$. For any $0 \leq \alpha \leq \mu^*$, we define the constrained SSR-optimal strategy $\mathcal{T}_\alpha$ as the strategy with maximum SSR among all strategies with return at least $\alpha$, i.e., $A(\mathcal{T}_\alpha) \geq \alpha$ and for all strategies $\mathcal{T}$ with $A(\mathcal{T} \geq \alpha)$, $\mathsf{S}(\mathcal{T}_\alpha) \geq \mathsf{S}(\mathcal{T})$. Note that while an SSR-optimal strategy can be chosen with one trading interval, a constrained SSR-optimal strategy may require more than one trading interval. We show that for some appropriate threshold $\alpha$, the constrained SSR-optimal strategy is a $\overline{\mathsf{Shrp}}_2$-optimal strategy.

**Proposition 4.6** *$\exists 0 \leq \alpha \leq \mu^*$ such that the constrained SSR-optimal strategy $\mathcal{T}_\alpha$ is $\overline{\mathsf{Shrp}}_2$-optimal.*

**Proof:** Let $\mathcal{T}$ be any $\overline{\mathsf{Shrp}}_2$-optimal strategy, and let $\alpha^* = A(\mathcal{T})$. Let $\mathcal{T}_{\alpha^*}$ be any constrained SSR-optimal strategy. Then $A(\mathcal{T}_{\alpha^*}) \geq A(\mathcal{T})$ and since $\mathsf{S}(\mathcal{T}_{\alpha^*}) \geq \mathsf{S}(\mathcal{T})$, we have that

$$0 \leq A(\mathcal{T}_{\alpha^*}) B(\mathcal{T}) - A(\mathcal{T}) B(\mathcal{T}_{\alpha^*}).$$

Suppose that $\overline{\mathsf{Shrp}}_2(\mathcal{T}_{\alpha^*}) < \overline{\mathsf{Shrp}}_2(\mathcal{T})$, then

$$0 \leq A(\mathcal{T}_{\alpha^*}) B(\mathcal{T}) - A(\mathcal{T}) B(\mathcal{T}_{\alpha^*}) < A(\mathcal{T}_{\alpha^*}) A(\mathcal{T}) \cdot (A(\mathcal{T}) - A(\mathcal{T}_{\alpha^*})).$$

Both $A(\mathcal{T}_{\alpha^*})$ and $A(\mathcal{T})$ are $> 0$, otherwise both strategies are inferior to $\mathcal{T}^*_\mu$; thus $A(\mathcal{T}) > A(\mathcal{T}_{\alpha^*})$, which is a contradiction. Therefore $\overline{\mathsf{Shrp}}_2(\mathcal{T}_{\alpha^*}) \geq \overline{\mathsf{Shrp}}_2(\mathcal{T})$ and so $\mathcal{T}_{\alpha^*}$ is $\overline{\mathsf{Shrp}}_2$-optimal. ∎

We will need the following technical property of any SSR-optimal interval (see appendix for proof).

**Proposition 4.7** *Let $J$ be a subinterval of an SSR-optimal interval $I$. Then, $\mu(J) \geq -f_{sp}$. Further, if $J$ is a prefix or suffix of $I$, then $\mu(J) > 0$.*

The next result establishes the intuitive result that adding an SSR-optimal interval to any trading strategy can only improve the strategy.

**Proposition 4.8** *Let $I_0$ be any SSR-optimal interval, and let $\mathcal{T}$ be any trading strategy. Let $\mathcal{T}' = I_0 \cup \mathcal{T}$. Then $A(\mathcal{T}') \geq A(\mathcal{T})$ and $\mathsf{S}(\mathcal{T}') \geq \mathsf{S}(\mathcal{T})$*

We can now give the intuition behind our algorithm. The starting point is Proposition 4.6, which says that it suffices to look for constrained SSR-optimal strategies. So the natural first choice is an unconstrained SSR-optimal interval $\mathcal{T}_0$. Either this will be $\overline{\mathsf{Shrp}}_2$ optimal or not. If not, it is because it has too small a return. So our next step is to add to this interval a new interval (possibly adjacent) with the property that the interval increases the return with *smallest* possible decrease in SSR, resulting in strategy $\mathcal{T}_1$. We repeat this process, constructing a sequence of trading strategies $\mathcal{T}_0, \mathcal{T}_1, \ldots$ with the property that $A(\mathcal{T}_i) > A(\mathcal{T}_{i-1})$, and among all other strategies $\mathcal{T}$ such that $A(\mathcal{T}) > A(\mathcal{T}_{i-1})$, $\mathsf{S}(\mathcal{T}_i) \geq \mathsf{S}(\mathcal{T})$. We then pick the strategy $\mathcal{T}_{i^*}$ with maximum $\overline{\mathsf{Shrp}}_2$ ratio among these strategies, which will be globally sharpe optimal.

Suppose that we have a current strategy, $\mathcal{T}_i$. We need to determine the next piece to add to this so that we increase the return, with smallest possible decrease in SSR. Let $\mathcal{T}_i$ be composed of the intervals $I_0, I_1, \ldots, I_d$. We replace each of these intervals by a special symbol, $, to signify that these regions are already included in the strategy. We thus obtain a *generalized returns sequence*, one in which some intervals are replaced by the $ symbol. A *generalized trading strategy* on the generalized return sequence must be composed of trades that do not contain the $ symbol. However trades may be adjacent to the $ symbol. A trade interval $I$ in a generalized trading strategy can be *isolated* (not adjacent to any $ symbol), *extending* (adjacent to one $ symbol), or *bridging* (adjacent to two $ symbols). In order to correctly account for the transactions cost, we need to change how we compute $A(I)$, so we introduce the new function $\bar{A}(I)$:

$$\bar{A}(I) = \begin{cases} A(I) & \text{I is isolated} \\ A(I) + \frac{f_{sp}}{n} & \text{I is extending} \\ A(I) + \frac{2f_{sp}}{n} & \text{I is bridging} \end{cases}$$

The *generalized simplified Sharp ratio (GSSR)* for generalized strategy $\mathcal{T} = \{I_1, \ldots, I_d\}$ is

$$\bar{\mathsf{S}}(\mathcal{T}) = \frac{\sum_{i=1\ldots d} \bar{A}(I_i)}{\sum_{i=1\ldots d} B(I_i)}$$

Similar to the notion of a maximal return optimal strategy, we introduce the notion of a maximal SSR-optimal (or GSSR-optimal) interval as one which cannot be extended in either direction without decreasing the SSR (or GSSR).

We now define generalized return sequences $\{R_0, R_1, \ldots\}$ as follows. $R_0$ is just the original returns sequence. Let $I_i$ be a maximal GSSR-optimal interval for $R_i$. We obtain the generalized sequence $R_{i+1}$ by replacing $I_i \subset R_i$ with the symbol $. We define any set of generalized sequences obtained in this way as *monotone*. We also refer to a member of a monotone set as monotone. Let $R_0, R_1, \ldots, R_k$ be a monotone sequence of gerenalized returns sequences, and let $I_0, I_1, \ldots, I_k$ be the maximal GSSR-optimal intervals corresponding to each sequence. By construction, $I_i$ is a maximal GSSR-optimal interval for $R_i$. We have defined $\bar{A}$ so that the SSR of the union of these

intervals in $R_0$ is given by

$$\mathsf{S}_{R_0}(I_0 \cup I_1 \cup \cdots \cup I_k) = \frac{\sum_{i=1}^d \bar{A}_{R_i}(I_i)}{\sum_{i=1}^d B(I_i)},$$

where the subscript $R_i$ indicates on which generalized return sequence the quantity is computed.

**Lemma 4.9** $\bar{\mathsf{S}}_{R_i}(I_i) \geq \bar{\mathsf{S}}_{R_{i+1}}(I_{i+1})$

**Proof:** Suppose that $\bar{\mathsf{S}}_{R_i}(I_i) < \bar{\mathsf{S}}_{R_{i+1}}(I_{i+1})$, and let $\$_i$ be the symbol that replaced $I_i$ in $R_i$ to obtain $R_{i+1}$. If $I_{i+1}$ is not adjacent with $\$_i$, then $I_i$ is not GSSR-optimal in $R_i$, a contradiction. If $I_{i+1}$ is adjacent with $\$_i$, then $I_i \cup I_{i+1}$ has higher GSSR (by Lemma 4.1), so once again $I_i$ is not GSSR-optimal in $R_i$. ∎

Now an easy induction, using Lemmas 4.1 and 4.9 gives,

**Corollary 4.10** $\mathsf{S}_{R_0}(I_0 \cup I_1 \cup \cdots \cup I_k) \geq \bar{\mathsf{S}}_{R_k}(I_k)$ for any $k$.

Analogous to Propositions 4.4, 4.7, 4.8, we have the following three propositions. Their proofs are almost identical, so we omit them.

**Proposition 4.11** *A GSSR-optimal strategy making one trade exists, and all maximal GSSR-optimal trades can be found in $O(N \log N)$ time.*

**Proposition 4.12** *Let $J$ be a subinterval of any GSSR-optimal interval $I$. Then $\mu(J) \geq -f_{sp}$. If $J$ is a prefix or suffix of $I$ that is not adjacent with the symbol "$\$$", then $\mu(J) > 0$.*

**Proposition 4.13** *Let $I_0$ be any GSSR-optimal interval, and let $\mathcal{T}$ be any generalized trading strategy. Let $\mathcal{T}' = I_0 \cup \mathcal{T}$. Then, $\bar{A}(\mathcal{T}') \geq \bar{A}(\mathcal{T})$ and $\bar{\mathsf{S}}(\mathcal{T}') \geq \bar{\mathsf{S}}(\mathcal{T})$.*

We now give the main result that will lead to the final algorithm to obtain the $\overline{\mathsf{Shrp}}_2$-optimal strategy. Its essential content is that given a monotone set of generalized returns sequences, $R_0, R_1, \ldots$, with corresponding GSSR-optimal intervals $I_0, I_1, \ldots$, for some $k$, $\mathcal{T} = I_0 \cup I_1 \cup \cdots \cup I_k$ is $\overline{\mathsf{Shrp}}_2$ optimal. We will need some preliminary results.

**Proposition 4.14** *For some $k$, $\mathcal{T}^* = I_0 \cup I_1 \cup \cdots \cup I_k$ is $\overline{\mathsf{Shrp}}_2$-optimal, where $I_i$ are the GSSR-optimal intervals corresponding to a monotone set of generalized returns sequences.*

**Proof:** First we show that there exists a $\overline{\mathsf{Shrp}}_2$-optimal strategy $\mathcal{T}_0$ that contains $I_0$. Indeed, let $\mathcal{T}$ be any $\overline{\mathsf{Shrp}}_2$-optimal strategy, and consider $\mathcal{T}_0 = \mathcal{I}_0 \cup \mathcal{T}$. By the Proposition 4.8, we have $\mathsf{S}(\mathcal{T}_0) \geq \mathsf{S}(\mathcal{T})$ and $A(\mathcal{T}_0) \geq A(\mathcal{T}) \geq 0$. Then,

$$\overline{\mathsf{Shrp}}_2(\mathcal{T}_0) - \overline{\mathsf{Shrp}}_2(\mathcal{T}) = \frac{A(\mathcal{T}_0)A(\mathcal{T})(A(\mathcal{T}_0) - A(\mathcal{T})) + B(\mathcal{T}_0)B(\mathcal{T})(\mathsf{S}(\mathcal{T}_0) - \mathsf{S}(\mathcal{T}))}{(B(\mathcal{T}_0) - A^2(\mathcal{T}_0))(B(\mathcal{T}) - A^2(\mathcal{T}))} \geq 0,$$



thus, $\mathcal{T}_0$ is $\overline{\mathsf{Shrp}}_2$-optimal.

Let $\mathcal{T}_k$ be a $\overline{\mathsf{Shrp}}_2$-optimal strategy that contains $I_0 \cup \cdots \cup I_k$. We know that $\mathcal{T}_0$ exists. If $\mathcal{T}_k = I_0 \cup \cdots \cup I_k$, then we are done. If, $\mathcal{T}_k = I_0 \cup \cdots \cup I_k \cup \mathcal{T}'$, with $\mathcal{T}' \neq \emptyset$, then we show that there must exist a $\overline{\mathsf{Shrp}}_2$-optimal strategy $\mathcal{T}_{k+1}$ which contains $I_0 \cup \cdots \cup I_{k+1}$, i.e., there is some other $\mathcal{T}'' \supseteq I_{k+1}$ such that $\mathcal{T}_{k+1} = I_0 \cup \cdots \cup I_k \cup \mathcal{T}''$ is $\overline{\mathsf{Shrp}}_2$-optimal. The proposition then follows by an easy induction.

Let $\mathcal{T}'' = \mathcal{T}' \cup I_{k+1}$. Then, $\bar{A}_{R_{k+1}}(\mathcal{T}'') \geq \bar{A}_{R_{k+1}}(\mathcal{T}')$ and $\bar{\mathsf{S}}_{R_{k+1}}(\mathcal{T}'') \geq \bar{\mathsf{S}}_{R_{k+1}}(\mathcal{T}')$ (Proposition 4.13). By Corollary 4.10 and the GSSR-optimality of $I_{k+1}$, we have that

$$\mathsf{S}(I_0 \cup \ldots \cup I_k) \geq \bar{\mathsf{S}}_{R_{k+1}}(I_{k+1}) \geq \bar{\mathsf{S}}_{R_{k+1}}(\mathcal{T}'') \geq \bar{\mathsf{S}}_{R_{k+1}}(\mathcal{T}')$$

From now on, we will drop the $R_{k+1}$ subscript. Let $A = A(I_0 \cup \ldots \cup I_k)$, $B = B(I_0 \cup \ldots \cup I_k)$, $A' = \bar{A}(\mathcal{T}')$, $B' = B(\mathcal{T}')$, $A'' = \bar{A}(\mathcal{T}'')$ and $B'' = B(\mathcal{T}'')$. Let $\overline{\mathsf{Shrp}}_2 = \overline{\mathsf{Shrp}}_2(I_0 \cup \ldots \cup I_k)$, $\overline{\mathsf{Shrp}}_2' = \overline{\mathsf{Shrp}}_2(I_0 \cup \ldots \cup I_k \cup \mathcal{T}')$ and $\overline{\mathsf{Shrp}}_2'' = \overline{\mathsf{Shrp}}_2(I_0 \cup \ldots \cup I_k \cup \mathcal{T}'')$. Thus,

$$\overline{\mathsf{Shrp}}_2 = \frac{A}{B - A^2}, \quad \overline{\mathsf{Shrp}}_2' = \frac{A + A'}{B + B' - (A + A')^2}, \quad \text{and} \quad \overline{\mathsf{Shrp}}_2'' = \frac{A + A''}{B + B'' - (A + A'')^2}.$$

Let $A'' = \alpha'' A$ and $A' = \alpha' A$, where $\alpha'' \geq \alpha' > 0$. Then, by direct computation, one obtains

$$\overline{\mathsf{Shrp}}_2' = \frac{A}{B + A^2 + \frac{\alpha'}{1+\alpha'}(\frac{B'}{\alpha'} - B - A^2)}, \quad \text{and} \quad \overline{\mathsf{Shrp}}_2'' = \frac{A}{B + A^2 + \frac{\alpha''}{1+\alpha''}(\frac{B''}{\alpha''} - B - A^2)},$$

Since $\overline{\mathsf{Shrp}}_2' > \overline{\mathsf{Shrp}}_2$, we conclude that $\frac{B'}{\alpha'} - B - A^2 < 0$. Since $\bar{\mathsf{S}}(\mathcal{T}'') \geq \bar{\mathsf{S}}(\mathcal{T}')$, we have that $\frac{B'}{\alpha'} \geq \frac{B''}{\alpha''}$, and since $\alpha'' \geq \alpha' > 0$, $\frac{\alpha''}{1+\alpha''} \geq \frac{\alpha'}{1+\alpha'} > 0$, therefore

$$\frac{\alpha''}{1+\alpha''}\left(\frac{B''}{\alpha''} - B - A^2\right) \leq \frac{\alpha'}{1+\alpha'}\left(\frac{B'}{\alpha'} - B - A^2\right) < 0,$$

and so $\overline{\mathsf{Shrp}}_2'' \geq \overline{\mathsf{Shrp}}_2'$, concluding the proof. ∎

By Proposition 4.14, a $\overline{\mathsf{Shrp}}_2$-optimal trading strategy can be obtained by constructing the strategies $\mathcal{T}_k$, and then picking the one with the maximum value for $\overline{\mathsf{Shrp}}_2$. The next proposition shows that this can be done in $O(n^2 \log n)$ time.

**Proposition 4.15** *A $\overline{\mathsf{Shrp}}_2$-optimal trading strategy can be found in time $O(n^2 \log n)$.*

**Proof:** $I_i$ can be obtained in $O(n \log n)$ time (Proposition 4.11). Since there are at most $n$ such intervals (since each must be non-empty), obtaining all the intervals is in $O(n^2 \log n)$.

Given the intervals, a single scan can be used to obtain the $k$ for which $\mathcal{T}_k$ is $\overline{\mathsf{Shrp}}_2$-optimal. ∎



One can improve the runtime to $O(n^2)$ if $O(n^2)$ memory is available, however, we do not discuss the details.

### 4.2.1 Approximation Ratio

We have given an algorithm that obtains a $\overline{\mathsf{Shrp}}_2$-optimal strategy. A modification to the algorithm constructs the hierarchy $\mathcal{T}_i$ and pick the one with the highest value of $\mathsf{Shrp}_2$. Suppose we have a $\overline{\mathsf{Shrp}}_2$-optimal strategy $\mathcal{T}$ and let $\mathcal{T}^*$ be a $\mathsf{Shrp}_2$-optimal strategy. Then by Proposition 4.6, it must be that $A(\mathcal{T}^*) > A(\mathcal{T})$ and that $\mathsf{S}(\mathcal{T}^*) \leq \mathsf{S}(\mathcal{T})$. Since $\overline{\mathsf{Shrp}}_2(\mathcal{T}) \geq \overline{\mathsf{Shrp}}_2(\mathcal{T}^*)$, we have that $A^*(B - A^2) - A(B^* - A^{*2}) \leq 0$, where $A = A(\mathcal{T}), A^* = A(\mathcal{T}^*), B = B(\mathcal{T}), B^* = B(\mathcal{T}^*)$. We can evaluate $\mathsf{Shrp}_2(\mathcal{T}^*) - \mathsf{Shrp}_2(\mathcal{T})$ to obtain

$$0 \leq \mathsf{Shrp}_2(\mathcal{T}^*) - \mathsf{Shrp}_2(\mathcal{T}) \leq \mathsf{Shrp}_2^* \cdot \frac{\frac{d}{n}f_{sp}^2}{\frac{d}{n}f_{sp}^2 + B - A^2}$$

When $B - A^2 = O(1)$, which is usually the case, we see that this is a very accurate approximation (since $f_{sp} \ll 1$).

## 4.3 Maximizing $\mathsf{Shrp}_1$

Once again, we introduce the slightly different quantity $\overline{\mathsf{Shrp}}_1$,

$$\mathsf{Shrp}_1(\mathcal{T}) = \frac{A(\mathcal{T})}{\sqrt{\frac{d}{n}f_{sp}^2 + B(\mathcal{T}) - A^2(\mathcal{T})}}, \qquad \overline{\mathsf{Shrp}}_1(\mathcal{T}) = \frac{A(\mathcal{T})}{\sqrt{B(\mathcal{T}) - A^2(\mathcal{T})}}.$$

We will optimize $\overline{\mathsf{Shrp}}_1(\mathcal{T})$. Since maximizing $\overline{\mathsf{Shrp}}_1(\mathcal{T})$ is equivalent to minimizing $1/\overline{\mathsf{Shrp}}_1^2(\mathcal{T})$ the problem reduces to maximizing

$$Q(\mathcal{T}) = \frac{A^2(\mathcal{T})}{B(T)}$$

The entire algorithm is analogous to that for maximizing $\overline{\mathsf{Shrp}}_2$ in the previous section, we only need to prove the analogs of Propositions 4.6 and 4.14.

**Proposition 4.16** $\exists 0 \leq \alpha \leq \mu^*$ such that the constrained SSR-optimal strategy $\mathcal{T}_\alpha$ is $Q$-optimal.

**Proof:** Let $\mathcal{T}$ be $\overline{\mathsf{Shrp}}_1$-optimal, and let $\alpha^* = A(\mathcal{T})$. Let $\mathcal{T}_{\alpha^*}$ be a corresponding constrained SSR-optimal strategy. $A(\mathcal{T}_{\alpha^*}) \geq A(\mathcal{T})$ and $\frac{A(\mathcal{T}_{\alpha^*})}{B(\mathcal{T}_{\alpha^*})} \geq \frac{A(\mathcal{T})}{B(\mathcal{T})}$. Multiplying these two inequalities gives that $\frac{A^2(\mathcal{T}_{\alpha^*})}{B(\mathcal{T}_{\alpha^*})} \geq \frac{A^2(\mathcal{T})}{B(\mathcal{T})}$, i.e. $\mathcal{T}_{\alpha^*}$ is also $Q$-optimal. ∎

**Proposition 4.17** For some $k$, $\mathcal{T}^* = I_0 \cup I_1 \cup \cdots \cup I_k$ is $Q$-optimal., where $I_i$ are the GSSR-optimal intervals corresponding to a monotone set of generalized returns sequences.



**Proof:** The proof is very similar to tho proof of Proposition 4.14. Let $\mathcal{T}$ be $Q$-optimal, and let $\mathcal{T}_0 = I_0 \cup \mathcal{T}$. Then $A(\mathcal{T}_0) \geq A(\mathcal{T})$ and $\mathsf{S}(\mathcal{T}_0) \geq \mathsf{S}(\mathcal{T})$. Multiplying these two inequalities give that $Q(\mathcal{T}_0) \geq Q(\mathcal{T})$, or that $\mathcal{T}_0$ is also $Q$-optimal.

Let $\mathcal{T}_k$ be a $Q$-optimal strategy that contains $I_0 \cup \cdots \cup I_k$. Introduce $\mathcal{T}', \mathcal{T}'' = I_{k+1} \cup \mathcal{T}'$ as in the proof of Proposition 4.14. Let $Q = Q(I_0 \cup \ldots \cup I_k)$, $Q' = Q(I_0 \cup \ldots \cup I_k \cup \mathcal{T}')$ and $Q'' = Q(I_0 \cup \ldots \cup I_k \cup \mathcal{T}'')$,

$$Q = \frac{A^2}{B}, \quad Q' = \frac{(A+A')^2}{B+B'}, \quad \text{and} \quad Q'' = \frac{(A+A'')^2}{B+B''}.$$

Following exactly the same logic as in the proof to Proposition 4.14, we only need to show that $Q'' \geq Q'$. Let $A'' = \alpha'' A$ and $A' = \alpha' A$, where $\alpha'' \geq \alpha' > 0$. $\frac{A''}{B''} \geq \frac{A'}{B'}$ implies that $\frac{B''}{\alpha''} \leq \frac{B'}{\alpha'}$, and so $\frac{B''}{\alpha''(\alpha''+2)} \leq \frac{B'}{\alpha'(\alpha'+1)}$. By direct computation, one obtains

$$Q' = \frac{A^2}{B + \left(1 - \frac{1}{(1+\alpha')^2}\right)\left(\frac{B'}{\alpha'(\alpha'+2)} - B\right)}, \quad Q'' = \frac{A^2}{B + \left(1 - \frac{1}{(1+\alpha'')^2}\right)\left(\frac{B''}{\alpha''(\alpha''+2)} - B\right)}.$$

Since $Q' > Q$, it must be that $\frac{B'}{\alpha'(\alpha'+2)} - B < 0$. Since $\alpha'' \geq \alpha'$, $1 - \frac{1}{(1+\alpha'')^2} \geq 1 - \frac{1}{(1+\alpha')^2}$, so we have that

$$\left(1 - \frac{1}{(1+\alpha'')^2}\right)\left(\frac{B''}{\alpha''(\alpha''+2)} - B\right) \leq \left(1 - \frac{1}{(1+\alpha')^2}\right)\left(\frac{B'}{\alpha'(\alpha'+2)} - B\right) < 0,$$

which implies that $Q'' \geq Q'$. ∎

### 4.3.1 Approximation Ratio

Once again, a modification to the algorithm constructs the hierarchy $\mathcal{T}_i$ and picks the one with the highest one with value of $\mathsf{Shrp}_1$. Suppose we have a $\overline{\mathsf{Shrp}}_1$-optimal strategy $\mathcal{T}$ and let $\mathcal{T}^*$ be a $\mathsf{Shrp}_1$-optimal strategy. By direct computation, and using the fact that $\overline{\mathsf{Shrp}}_1(\mathcal{T}) \geq \overline{\mathsf{Shrp}}_1(\mathcal{T}^*) \implies A^{*2}B - A^2B^* \leq 0$, we get

$$0 \leq \mathsf{Shrp}_1^2(\mathcal{T}^*) - \mathsf{Shrp}_1^2(\mathcal{T}) \leq \mathsf{Shrp}_1^2(\mathcal{T}^*) \frac{\frac{df_{sp}^2}{n}}{\frac{df_{sp}^2}{n} + B}$$

which gives an approximation ratio of $\sqrt{1 - O(f_{sp}^2)}$ when $B = O(1)$.



## 5 Discussion

Our main goal was to provide the theoretical basis for the computation of *a posteriori* optimal trading strategies, with respect to various criteria. In particular, we have presented the algorithms, together with the proofs of their correctness. The highlights of our contributions are that return and sterling optimal strategies can be computed very efficiently, even with constraints on the number of trades. Sharpe optimal strategies prove to be much tougher to compute. However, for slightly modified notions of the Sharpe ratio, where one ignores the impact of bid-ask spread squared we can compute the optimal strategy efficiently. This is a reasonable approach since in most cases, the bid-ask spread is $\sim 10^{-4}$. We also show that this modified optimal strategy is not far from optimal with respect to the unmodified Sharpe ratio.

Interesting topics of future research are to use these optimal strategies to learn how to trade optimally, which was the original motivation of this work. Further, one could use them to benchmark trading strategies as well as markets. A natural open problem is whether Sharpe optimal strategies can be computed under constraints on the number of trades. We suspect that the monotone hierarchy we created with respect to the SSR has an important role to play, but the result has been elusive.

The algorithms presented here can be useful if one could relax Assumption **A1**. We have used such an assumption because it considerably simplifies the analysis, and in many cases, the optimal strategy is in fact an all-or-nothing strategy.

At the core of some of our algorithms is a new technique for optimizing quotients over intervals of a sequence. This technique is based on relating the problem to convex set operations, and for our purposes has direct application to optimizing the $MDD$, the simplified Sharpe ratio (SSR), which is an integral component in optimizing the Sharpe ratio, and the Downside Deviation Ratio (DDR). This technique may be of more general use in optimizing other financially important criteria. As a result, such an algorithm may be of independent interest.

## References


[1] Online encyclopedia of financial terms. *http://www.investopedia.com*, 2004.

[2] Arjan Berkelaar and Roy Kouwenberg. Dynamic asset allocation and downside-risk aversion. citeseer.ist.psu.edu/berkelaar00dynamic.html.

[3] Tomasz R. Bielecki, Jean-Philippe Chancelier, Stanley Pliska, and Agnes Sulem. Risk sensitive portfolio optimization with transaction costs. *To appear*, 2004.





[4] Tomasz R. Bielecki, Stanley Pliska, and Jiongmin Yong. Optimal investment decisions for a portfolio with a rolling horizon bond and a discount bond. *To appear*, 2004.

[5] Timothy M. Chan. Output-sensitive results on convex hulls, extreme points, and related problems. In *Proc. 11th Symposium on Computational Geometry*, pages 10 – 19, 1995.

[6] Thomas H. Cormen, Charles E. Leiserson, Ronald L. Rivest, and Clifford Stein. *Introduction to Algorithms*. Mcgraw-Hill, 2001.

[7] Sergei Issaenko. Domenico Cuoco, Hua He. Optimal dynamic trading strategies with risk limits. *SSRN Electronic Paper Collection*, 2001.

[8] Bhaswar Gupta and Manolis Chatiras. The interdependence of managed futures risk measures. pages 203–220. Center for International Securities and Derivatives Markets, October 2003.

[9] Matthew Saffell. John Moody. Learning to trade via direct reinforcement. *IEEE Transactions on Neural Networks*, 12(4):875–889, 2001.

[10] Hong Liu. Optimal consumption and investment with transaction costs and multiple risky assets. *The Journal of Finance*, 59(1):289, 2004.

[11] Stephen Boyd Miguel Sousa Lobo, Maryam Fazel. Portfolio optimization with linear and fixed transaction costs. *Annals of Operations Research, special issue on financial optimization, to appear*, 2006.

[12] Oliver Mihatsch and Ralph Neuneier. Risk-sensitive reinforcement learning. *Mach. Learn.*, 49(2-3):267–290, 2002.

[13] Jiming Liu. Samuel P. M. Choi. Optimal time-constrained trading strategies for autonomous agents. *In Proceedings of International ICSC Symposium on Multi-agents and Mobile Agents in Virtual Organizations and E-Commerce (MAMA 2000)*, 2000.

[14] G. W. P. Thompson. Optimal trading of an asset driven by a hidden markov process in the presence of fixed transaction cost. In *Collection of research papers*. Judge Institute of Management Working Papers, 2002.

[15] Jiqing Han. Yang Liu, Xiaohui Yu. Sharp ratio-oriented active trading: A learning approach. *SSRN Electronic Paper Collection*, 2001.

[16] Valeri I. Zakamouline. Optimal portfolio selection with both fixed and proportional transaction costs for a CRRA investor with finite horizon. Norwegian School of Economics and Business Administration, 2002.




# A  Technical Proofs

**Lemma 3.13.**

**Proof:** Let $A = O(m)$ be the size of the current convex hull. For (1), we do not change the points at all, we simply compute the new tangent point for $\mathbf{p}' = \mathbf{p}+\mathbf{v}$, which can be accomplished in $O(\log A)$. (2) is equivalent to shifting $\mathbf{p}$ by $-\mathbf{v}$. To prove (3), notice that if we remove $(d_0, c_0)$, then the new leftmost point becomes $(d_1, c_1)$ and we immediately have the new convex hull $\mathsf{nxt}(1), \mathsf{nxt}(\mathsf{nxt}(1)), \ldots$. Thus we can find the new upper tangent point in $O(\log A') = O(\log m)$, where $A'$ is the size of the new convex hull. Further, deleting $(d_0, c_0)$ requires first deleting the backward pointers of the points that it points to $O(\log A)$, and then deleting the point itself, and its forward pointers (it has no backward pointers), $O(\log A)$. To prove (4), note that when we add $(d_{-1}, c_{-1})$, $\mathsf{nxt}(-1)$ is exactly the upper tangent point from $\mathbf{p}' = (d_{-1}, c_{-1})$ to the current convex hull. This can be computed in $O(\log A)$. We now need to add all the necessary pointers into the data structure. For each forward pointer we add, we will add the coresponding backward pointer as well. We need a pointer at position $2^j$ in the convex hull of $(d_{-1}, c_{-1})$. But this is exactly the point at position $2^j - 1$ in the convex hull of point $\mathsf{nxt}(-1)$. Since $\mathsf{nxt}(-1)$ maintains a pointer to point $2^j$ in its convex hull, and this point will have a backward pointer by one step of this same convex hull, we can construct the forward an backward pointer for point $2^j$ in the convex hull of $(d_{-1}, c_{-1})$ in constant time, requiring total time $O(\log A') = O(\log m)$ to construct all the new forward and backward pointers, where $A'$ is the size of the new convex hull. We now construct the new upper tangent point from $\mathbf{p}$ to the new convex hull of $(d_{-1}, c_{-1})$ in $O(\log A')$ time. The entire process is therefore $O(\log m)$. ∎

The algorithm that we have just described is a general purpose algorithm for efficiently maintaining the upper tangent point to *any* set of points, as long as only a limited set of operations is allowed on the set of points and the source point.

**Proposition 3.18.**

**Proof:** First we show that for every trading interval $I^* = [t_a, t_b]$ in $\mathcal{T}_\mu^*$ with $I \cap I^* \neq \emptyset$, one can pick $\mathcal{T}_{\mathsf{Strl}}^K$ such that $I^* \subseteq I$. Suppose to the contrary, that for some $I^*$, either $t_a < t_l$ and $t_b \geq t_l$ or $t_a \leq t_r$ and $t_b > t_r$. We will extend $I$ without decreasing the Sterling ratio of $\mathcal{T}_{\mathsf{Strl}}^K$ so that $I^* \subseteq I$. Suppose $t_a < t_l$ and $t_b \geq t_l$ (a similar argument holds for $t_a \leq t_r$ and $t_b > t_r$). There are two cases:

i. $I^*$ does not intersects any other interval of $\mathcal{T}_{\mathsf{Strl}}^K$: Applying Lemma 3.17 to $I^*$, we have: $C_a \leq C_l$. Thus by extending $I$ to $[t_a, t_r]$, the return of the interval cannot decrease. Since $MDD(I^*) \leq f_{sp}$, this extension cannot increase the $MDD(\mathcal{T}_{\mathsf{Strl}}^K)$, since we already have that $MDD(\mathcal{T}_{\mathsf{Strl}}^K) \geq f_{sp}$.



ii. $I^*$ intersects with the previous trading interval of strategy $\mathcal{T}_{\text{Strl}}^K$: $\exists I' = [t_{l'}, t_{r'}] \in \mathcal{T}_{\text{Strl}}^K$ such that $t_a \leq t_{r'} < t_l$. Since $[t_{r'+1}, t_{l-1}]$ is a subinterval of $I^*$, $\sum_{j=r'+1}^{l-1} \hat{s}_j \geq -f_{sp}$ (Lemma 3.16). If we merge $I$ and $I'$ by adding the interval $[t_{r'+1}, t_{l-1}]$ into $\mathcal{T}_{\text{Strl}}^K$, we save on the transaction cost of $f_{sp}$, and so the total return will not decrease. We show that the $MDD$ has not increased. Since $C_{r'}$ is a maximum in $[t_{l'}, t_{r'}]$, the drawdown for all points in $[t_{r'+1}, t_l]$ is at most $f_{sp}$. Since $C_l$ is a minimum in $[t_l, t_r]$, we conclude that the drawdown for any point in $[t_l, t_r]$ is at most $\max\{f_{sp}, MDD(I)\}$. Since $MDD(\mathcal{T}_{\text{Strl}}^K) \geq f_{sp}$, we conclude that this merger does not increase the $MDD$.

Note that $\mu(I) \geq 0$ otherwise we improve the return of $\mathcal{T}_{\text{Strl}}^K$ by removing I, without increasing the $MDD$, ans so $\mathcal{T}_{\text{Strl}}^K$ cannot possibly be optimal. Thus, without loss of generality, we assume that the return of $I$ is positive. Suppose that $I \cap \mathcal{T}_\mu^* = \emptyset$. Then, by adding $I$ to $\mathcal{T}_\mu^*$, we strictly increase the return, a contradiction on the optimality of $\mathcal{T}_\mu^*$. Thus, every interval of $\mathcal{T}_{\text{Strl}}^K$ contains at least one interval of $\mathcal{T}_\mu^*$. Now consider the maximal prefix $P_{\max}$ of $I$ that does not overlap with any interval of $\mathcal{T}_\mu^*$. Since we know that $I$ contains some interval of $\mathcal{T}_\mu^*$, we conclude that this maximal prefix must be adjacent to some interval of $\mathcal{T}_\mu^*$. By Lemma 3.16, this interval has strictly negative return, so removing it from $I$ strictly increase the return of $\mathcal{T}_{\text{Strl}}^K$, without increasing its $MDD$. This contradicts the optimality of $\mathcal{T}_{\text{Strl}}^K$, thus, $P_{\max}$ must be empty. Similarily, the maximal suffix of $I$ that is non-intersecting with $\mathcal{T}_\mu^*$ must be empty, concluding the proof. ∎

**Lemma 3.19.**

**Proof:** If $K = K_0$, then $\mathcal{T}_\mu^*$ itself is K-Sterling-optimal. If $K < K_0$, we show that if the number of trades made is less than $K$, we can always add one more interval without decreasing the Sterling ratio of the strategy. First, note that $\mathcal{T}_{\text{Strl}}^K$ cannot contain all the intervals of $\mathcal{T}_\mu^*$, as otherwise (by the pigeonhole principle) at least one interval $I = [t_l, t_r]$ of $\mathcal{T}_{\text{Strl}}^K$ contains two consecutive intervals $I_1 = [t_{l_1}, t_{r_1}]$ and $I_1 = [t_{l_2}, t_{r_2}]$ of $\mathcal{T}_\mu^*$. The region between these two intervals has return less than $-f_{sp}$ (Lemma 3.16), so breaking up $I$ into the two intervals $[t_l, t_{r_1}]$ and $[t_{l_2}, t_r]$ will strictly increase the return, without increasing the $MDD$, contradicting the optimality of $\mathcal{T}_{\text{Strl}}^K$. If $\mathcal{T}_{\text{Strl}}^K$ does not contain some interval of $\mathcal{T}_\mu^*$, then by adding this interval, we do not decrease the return or the $MDD$ (Lemma 3.16), since the $MDD$ is already $\geq f_{sp}$. ∎

**Proposition 4.7.**

**Proof:** If $J$ is a prefix or suffix of $I$ and $\mu(J) \leq 0$, then deleting $J$ from $I$ gives at least as much return, with smaller sum of squared returns, contradicting the SSR-optimality of $I$. Suppose that $I = L \cup J \cup R$ where $L$ and $R$ are nonempty subintervals of $I$. If $\frac{\mu(J)+\mu(R)}{s^2(J)+s^2(R)} < \frac{-f_{sp}+\mu(L)}{s^2(L)}$, then by



Lemma 4.1,
$$\mathsf{S}(I) = \frac{-f_{sp} + \mu(L) + \mu(J) + \mu(R)}{s^2(L) + s^2(J) + s^2(R)} < \frac{-f_{sp} + \mu(L)}{s^2(L)} = \mathsf{S}(L) \tag{*}$$

This contradicts the optimality of $I$, so we have
$$\frac{\mu(J) + \mu(R)}{s^2(J) + s^2(R)} \geq \frac{-f_{sp} + \mu(L)}{s^2(L)}.$$

Now, suppose that $\mu(J) < -f_{sp}$. Using e̊q:* and Lemma 4.1, we find that
$$\mathsf{S}(I) \leq \frac{\mu(J) + \mu(R)}{s^2(J) + s^2(R)} < \frac{-f_{sp} + \mu(R)}{s^2(J) + s^2(R)} < \frac{-f_{sp} + \mu(R)}{s^2(R)} = \mathsf{S}(R),$$

because $s^2(J) > 0$. This contradicts the SSR-optimality of $I$, so $\mu(J) \geq -f_{sp}$. ∎

**Proposition 4.8.**

**Proof:** If $I_0$ is contained in some interval of $\mathcal{T}$, then there is nothing to prove, so suppose that $\mathcal{T} \cup I_0 \neq \mathcal{T}$, and that $\mathcal{T}$ contains $d \geq 1$ trades $I_1, \ldots, I_d$. Note that $\mathsf{S}(\mathcal{T}) = A(\mathcal{T})/B(\mathcal{T})$. If $I_0$ and $\mathcal{T}$ do not intersect, then $\mathcal{T}' = I_0 \cup \mathcal{T} = \{I_0, I_1, \ldots, I_d\}$. $A(\mathcal{T}') = A(\mathcal{T}) + A(I_0) \geq A(\mathcal{T})$, because $A(I_0) \geq 0$. Since $I_0$ is SSR-optimal, $\mathsf{S}(I_0) = \frac{A(I_0)}{B(I_0)} \geq \frac{A(\mathcal{T})}{B(\mathcal{T})} = \mathsf{S}(\mathcal{T})$, so by lemma 4.1,
$$\mathsf{S}(\mathcal{T}') = \frac{A(\mathcal{T}')}{B(\mathcal{T}')} = \frac{A(\mathcal{T}) + A(I_0)}{B(\mathcal{T}) + B(I_0)} \geq \frac{A(\mathcal{T})}{B(\mathcal{T})} = \mathsf{S}(\mathcal{T}).$$

Suppose that $\mathcal{T} \cap I_0 \neq \emptyset$. We can decompose $\mathcal{T}$ into four parts (each part could be empty): $\mathcal{T} = S_1 \cup S_2 \cup I_l \cup I_r$, where $S_1$ contains intervals that do not intersect with $I_0$, $S_2$ contains intervals that are contained in $I_0$, $I_l$ is not contained in $I_0$ but overlaps $I_0$ on the left, and $I_r$ is not contained in $I_0$ but overlaps $I_0$ on the right. $\mathcal{T}' = I_0 \cup \mathcal{T} = S_1 \cup I_l \cup I_0 \cup I_r$, i.e., adding $I_0$ combines all the trades in $\{S_2, I_l, I_r\}$ into one trade. Since the internal regions of $I_0$ have return at least $-f_{sp}$ and any prefix and suffix of $I_0$ has positive return (Proposition 4.7), we see that merging any two consecutive trades overlapping $I_0$ decreases the number of trades by one, hence increases the return by $f_{sp}$ and the added interval loses at most $f_{sp}$, hence this merge can only increase $A(\mathcal{T})$. If either $I_l$ or $I_r$ are empty, then we are addionally adding a prefix or suffix of $I_0$ without changing the number of trades, which also increases $A(\mathcal{T})$, thus we see that $A(\mathcal{T}') \geq A(\mathcal{T})$.



Let's introduce the following definitions,

$$\begin{aligned}
A_1 &= A(S_1) + \frac{1}{n}(\mu(I_l \cap \overline{I_0}) + \mu(I_r \cap \overline{I_0})) \\
A_2 &= A(S_2) + \frac{1}{n}(\mu(I_l \cap I_0) - f_{sp} + \mu(I_r \cap I_0) - f_{sp}) \\
B_1 &= B(S_1) + \frac{1}{n}(s^2(I_l \cap \overline{I_0}) + s^2(I_r \cap \overline{I_0})) \\
B_2 &= B(S_2) + \frac{1}{n}(s^2(I_l \cap I_0) + s^2(I_r \cap I_0)),
\end{aligned}$$

where $\overline{I_0}$ is the complement of $I_0$. Letting $A_0 = A(I_0)$ and $B_0 = B(I_0)$, we then have

$$\mathsf{S}(\mathcal{T}) = \frac{A_1 + A_2}{B_1 + B_2}, \text{ and } \mathsf{S}(\mathcal{T}') = \frac{A_1 + A_0}{B_1 + B_0}.$$

Note that $\mathsf{S}(S_2 \cup (I_l \cap \overline{I_0}) \cup (I_r \cap \overline{I_0})) = \frac{A_2}{B_2}$, so by the optimality of $I_0$, $\frac{A_2}{B_2} \leq \frac{A_0}{B_0}$. We show that

$$\frac{\frac{1}{n}\mu(I_l \cap \overline{I_0})}{\frac{1}{n}s^2(I_l \cap \overline{I_0})} \leq \frac{A_0}{B_0}, \text{ and } \frac{\frac{1}{n}\mu(I_r \cap \overline{I_0})}{\frac{1}{n}s^2(I_r \cap \overline{I_0})} \leq \frac{A_0}{B_0}. \tag{**}$$

If not, then suppose (for example) that $\frac{\frac{1}{n}\mu(I_l \cap \overline{I_0})}{\frac{1}{n}s^2(I_l \cap \overline{I_0})} > \frac{A_0}{B_0}$. Then,

$$\mathsf{S}(I_0 \cup I_l) = \frac{A_0 + \frac{1}{n}\mu(I_l \cap \overline{I_0})}{B_0 + \frac{1}{n}s^2(I_l \cap \overline{I_0})} > \frac{A_0}{B_0} = \mathsf{S}(I_0)$$

contradicting the SSR-optimality of $I_0$. Again, by the optimality of $I_0$, $\mathsf{S}(S_1) = \frac{A(S_1)}{B(S_2)} \leq \frac{A_0}{B_0} = \mathsf{S}(I_0)$. it also has to be sharper than strategy $S_1$. Thus, using (**) and Lemma 4.1 we have that $\frac{A_1}{B_1} \leq \frac{A_0}{B_0}$. Because $A_2$ is obtained from the returns of a collection of subintervals of $I_0$, it follows from Proposition 4.7 that $A_2 \leq A_0$. Now suppose that $\mathsf{S}(\mathcal{T}) > \mathsf{S}(\mathcal{T}')$, i.e.,

$$(A_1 + A_2)(B_1 + B_0) - (A_1 + A_0)(B_1 + B_2) > 0.$$

Since $A_2 \leq A_0$, it follows that $B_2 \leq B_0$. Rearranging terms in the equation above, we have that

$$\begin{aligned}
\frac{A_2 B_1 - A_1 B_2}{B_0(B_1 + B_2)} &> \frac{A_0}{B_0} - \frac{A_1 + A_2}{B_1 + B_2}, \\
&\geq \frac{A_2}{B_2} - \frac{A_1 + A_2}{B_1 + B_2} = \frac{A_2 B_1 - A_1 B_2}{B_2(B_1 + B_2)}.
\end{aligned}$$

Since $\mathsf{S}(I_0) \geq \mathsf{S}(\mathcal{T})$, the first inequality shows that $A_2 B_1 - A_1 B_2 > 0$. The second inequality then implies that $B_2 > B_0$, a contradiction. ∎

45